\documentclass[12pt,prl,nofootinbib]{revtex4}
\usepackage[utf8]{inputenc}
\usepackage{amsmath}
\usepackage{amsfonts}
\usepackage{amssymb}
\usepackage{graphicx}
\usepackage{grffile}
\usepackage{color}
\usepackage{xcolor}
\usepackage{textgreek}
\usepackage{hyperref}
\usepackage{multirow}
\usepackage[normalem]{ulem}
\usepackage{color}
\usepackage{natbib}
\definecolor{light}{rgb}{0.5, 0.5, 0.5}

\date{\today}
\begin{document}
\title{Controlling the morphologies and dynamics in three-dimensional tissues}

\author{Rajsekhar Das$^1$ }
\affiliation{Department of Chemistry, University of Texas at Austin, Austin, Texas 78712, USA}
\author{Xin Li$^1$}
\affiliation{Department of Chemistry, University of Texas at Austin, Austin, Texas 78712, USA}
\author{Sumit Sinha$^2$}
\affiliation{Department of Physics, University of Texas at Austin, Austin, Texas 78712, USA}
\author{D. Thirumalai$^{1,2}$}
\affiliation{$^1$Department of Chemistry, University of Texas at Austin, Austin, Texas 78712, USA}
\affiliation{$^2$Department of Physics, University of Texas at Austin, Austin, Texas 78712, USA}
\begin{abstract}
A number of factors, such as, cell-cell interactions and self-propulsion of cells driven by cytoskeletal forces determine tissue morphologies and dynamics. To explore the interplay between these factors in controlling the dynamics at the tissue scale, we created a minimal three dimensional model in which  short-range repulsive elastic forces account for cell-cell  interactions. Self-propulsion is modeled as active uncorrelated random stochastic forces, with strength $\mu$, that act on individual cells and is the only source of cell motility. Strikingly, variations in polydispersity in cell sizes ($\Sigma$) and cell elasticity ($E$), results in the formation of a variety of distinct ``phases", driven entirely by $\mu$. At low $E$, the tissue behaves like a liquid, at all values of $\Sigma$, whereas at high $E$ and $\Sigma$, it has the characteristics of a glass. The tissue crystallizes at low $\Sigma$ provided $E$ exceeds a critical value.  Over a narrow range of $E$ and $\Sigma$, that lies between the boundaries of the liquid and glass phase, the effective viscosity increases like in a glass as the  cell density increases and saturates as the cells are compressed beyond a certain value, creating the viscosity saturation (VS) phase. The VS phase does not form in systems at finite temperature in which the dynamics satisfies the Fluctuation Dissipation Theorem.  In the glass phase, the tissue  exhibits aging (relaxation times depend on the waiting time) behavior at high $E$ values. Our findings provide a framework for designing tissues with tunable material properties by controlling the physical characteristics of cells.      

\end{abstract}
\maketitle 
\section{Introduction}
Mechanical and morphological properties of cells are of fundamental importance to embryogenesis and cancer progression~\cite{DonaldIngber2003,Guilak2009,Tambe2011,Kumar2009,petridou2019tissue,PETRIDOU20211914}. The material properties of tissues change so dramatically, especially during development, that they are reminiscent of phase transitions~\cite{Huebner2018}. Indeed,  the sharp changes in tissue connectivity found during morphogenesis have been analyzed using the language and concepts familiar in the theory of phase transitions~\cite{MORITA2017354,Mongera2018,Barriga2018,Petridou2019,li2024emergence,PETRIDOU20211914}. Strikingly, the equilibrium rigidity percolation theory~\cite{Thorpe1983,Jacobs1995PRL,Jacobs1996PRE}, intended to describe network glasses, was recently used to quantitatively accounts for zebrafish morphogenesis~\cite{PETRIDOU20211914}. 

Not surprisingly, the morphological changes are also accompanied by substantial variations in  the collective cell dynamics.  Cells  exhibit a broad range of dynamics spanning  glass~\cite{Schötz2013,Angelini2011,Nnetu2012,malmi2018cell},  solid~\cite{fung1993biomechanics,Vandiver2009}, and liquid like behavior~\cite{Liquidliketissue,Beysens2000,ranft2010fluidization,matoz2017cell,malmi2018cell}. A number of factors control the observed range of dynamical behavior. For instance,  homeostasis between cell division and apoptosis fluidizes tissues at long times, which was first shown in a pioneering study \cite{ranft2010fluidization}. In contrast, imaging experiments found glass-like dynamics,  characterized by the vanishing of the migration speed in confluent epithelial cells at high cell densities in two dimensions (2D) \cite{Angelini2011}. Notably, the velocity fields were spatially heterogeneous, varying from region to region (sub-sample to sub-sample like in glasses~\cite{Thirumalai89PRA}). Despite the crucial difference between synthetic materials and cellular systems in which cell division plays an important role in determining the dynamics~\cite{Angelini2011}, experimental findings in the latter are reminiscent of dynamical heterogeneity in glass forming materials~\cite{Biroli13JCP,Kirkpatrick15RMP}. Indeed, simulations using vertex models showed that there is a transition from glass to fluid~\cite{bi2016motility} in 2D confluent tissues, driven by self-propulsion (active force). Finally, theory and simulations accounting for cell division and apoptosis, and mechanical feedback exhibit complex dynamical behavior~\cite{malmi2018cell}. Variations in cell division rate and the strength of mechanical feedback result in sub-diffusive, diffusive, and even hyper diffusive dynamics~\cite{Sinha25SoftMatter}. 

An important issue that has received less attention is how changes in tissue morphology, which can be controlled by varying cell softness, affect collective dynamics, in non-confluent tissues in which cell division plays an insignificant role.  A previous insightful study~\cite{PETRIDOU20211914} has provided a link between morphology describing geometric networks in zebrafish blastoderm and the dynamics at a fixed value of cell elasticity. The observed morphological changes were quantitatively described using scaling laws that are valid near (second order) phase transitions~\cite{Thorpe1983,Jacobs1995PRL,Jacobs1996PRE}. In particular, it was shown that the non-confluent tissue undergoes a rigidity percolation transition \cite{Jacobs1995PRL} that is characterized by an abrupt increase in the cell-cell connectivity. Strikingly, the geometric changes are also reflected in viscosity (a collective property), which exhibits an unusual behavior as a function of cell packing fraction, $\phi$ ~\cite{PETRIDOU20211914}. At $\phi \le \phi_S$ the viscosity increases sharply, like in a conventional glassy materials~\cite{ANGELL199113}, but saturates when $\phi$ increases beyond a critical value, $\phi_S$,  a behavior that is rarely observed in synthetic materials \cite{sumitCommentary}. We refer to this as the viscosity saturation (VS) regime.    Computer simulations~\cite{Das2023}, using a particle-based model of a two-dimensional (2D) tissue at a fixed value of the cell softness (elasticity), found that the plateau in $\eta$ above $\phi_S$ is a consequence of saturation in the available free area per cell beyond $\phi_S$. If the value of the elasticity is chosen such that the cells interpenetrate to some extent then $\eta$ increases rapidly when $\phi \leq \phi_S$ and saturates beyond $\phi_S$ provided there is polydispersity ($\Sigma$) in the cell size. 

The applicability of the rigidity percolation theory  \cite{PETRIDOU20211914} was rationalized by purely a geometric criterion associated with cell-cell contact topology, which is surprising given the complexity of morphogenesis.  However, it is unclear how the network of cellular contacts would change if the cell elasticity is altered. This is not merely problem of an academic interest because cell softness plays a critical role in many biological properties, such as cell motility and cancer progression~\cite{Luo2016,Cross2007}.
For instance, metastatic cancer cells are softer than normal cells~\cite{Cross2007,Alibert17BiolCell}, and  exhibit increased migration~\cite{Liu2020,Watanabe2014,Gensbittel21DevCell}. Other experiments~\cite{Zhang2014,harn2015,rianna2020direct,Gensbittel21DevCell} also suggested that higher cell motility is usually correlated with lower stiffness because it helps the cells to deform more readily, allowing them to squeeze through constricted spaces during migration or invasion.  
The cell elastic modulus \textcolor{black}{($E$)} could vary from $1.4\times 10^{-4}MPa$ (human melanoma WM115 cells~\cite{Weder2014}) to $1.6\times10^{-3}MPa$ (non-malignant bladder HCV29 cells~\cite{Ramos2014,gnanachandran2022discriminating})  depending on the cell type. How the order of magnitude change  in cell elasticity impacts the morphology and collective dynamics in non-confluent tissue in three dimension (3D) is unknown. 

Here, we explore the morphological changes and the associated impact on the dynamics that occur as the cell softness and polydispersity are varied using a minimal computational model in which short range elastic (Hertz) forces are combined with stochastic self-propulsion forces. Although we are motivated by biological considerations, the work provides a framework for understanding the connections between  material properties and dynamics in active soft matter systems. The central results, which were obtained using 3D simulations are: (i) For a low value of cell softness, the cells behave like a fluid independent of the cell size variations. (ii) Conversely, at high values of cell elasticity, the tissue exhibits characteristics of a glass provided polydispersity ($\Sigma$) in the cell sizes is modest. (iii) In contrast, the tissue has crystal-like structures at low  $\Sigma$. (iv) At intermediate values of cell softness, the tissue exhibits viscosity increases as the volume fraction ($\phi$) increases, which follows the Vogel-Fulcher-Tamman (VFT) relation~\cite{Fulcher,Tammann} as long as $\phi$ is less than a critical critical value, $\phi_S$. When  $\phi$ exceeds $\phi_S$, the viscosity saturates. 
(v) The range of dynamics and the associated  morphologies require self-propulsion forces, and cannot be captured by varying temperature.

\section*{Results}
\label{Results}

\begin{figure}[!htpb]
\begin{center}
\includegraphics[width=0.98\columnwidth]{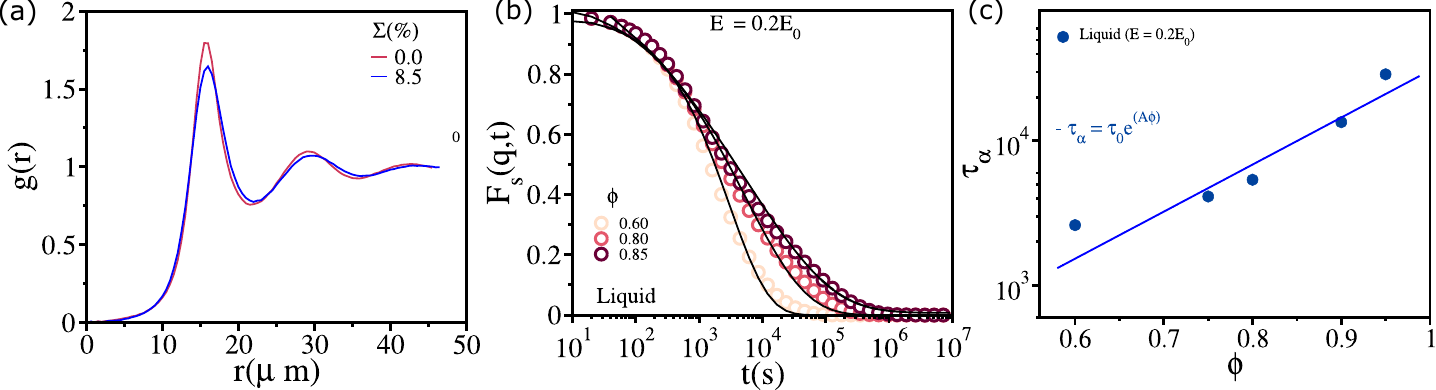}\caption{\textbf{Liquid states:} (a) The pair-correlation function $g(r)$ as a function of $r$ for $\Sigma =0.0$ (red) at $E = 0.1E_0$ and $\Sigma =8.5$ and $E =0.2E_0$ (blue)  at $\phi=0.85$. (b) Self-intermediate scattering function, $F_s(q,t)$, as a function of $t$ for $0.60 \leq \phi \leq 0.85$ for $E = 0.2E_0$ \textcolor{black}{($E_0\equiv 10^{-3}$MPa\cite{malmi2018cell})}. The black solid lines are fits to $F_s(q,t) = \tau_0\exp[-(t/\tau)^{\beta}]$. (c) $\tau_\alpha$ as a function of $\phi$ for $E=0.2E_0$. The solid line is the Arrhenius fit ($\tau_\alpha = \tau_0\exp(A\phi)$). For (b) and (c) polydispersity $\Sigma =8.5\%$. The value of $A$  is $A \sim 7.5$.}    
\label{Fig2}
\end{center}
\end{figure}
\textbf{Fluid-like states:}
We characterize the dynamics using the self-intermediate scattering function,
\begin{equation}
    F_s(q,t) = \frac{1}{N}\left\langle\sum_{j =1}^{N}\exp[ -i\vec{q}\cdot(\vec{r}_j(t)-\vec{r}_j(0))]\right\rangle,
    \label{fsqt}
\end{equation}
where the wave vector  $q = 2\pi/R_{\text{max}}$, and $R_{\text{max}}$ is the location of the first peak in the pair correlation function $g(r)$, (see Fig.~\ref{Fig2} (a) and Fig.~\ref{structure} (a)), and $\vec{r}_j(t)$ is the position of the $j^{th}$ cell at time $t$. The pair correlation function is given by 
\begin{equation}
    g(r) = \frac{1}{\rho}\left\langle \frac{1}{N}\sum_i^N\sum_{j\neq i}^N \delta\left(r - |\vec{r}_i - \vec{r}_j |\right)\right\rangle,
    \label{gr}
\end{equation}
where $\rho = \tfrac{N}{L^3}$ is the number density, $\delta$ is the Dirac delta function, $\vec{r}_i$ is the position of the $i^{th}$ cell, and the angular bracket $\left\langle\right\rangle$ is an ensemble average. The relaxation time $\tau_\alpha$ is taken to be as the time at which 
\begin{equation}
    F_s(q,t = \tau_\alpha) = 1/e.
    \label{tauAlpha}
\end{equation} 

The pair-correlation function $g(r)$ (Eqn.~\eqref{gr}) shows that there is no long-range order (Fig.~\ref{Fig2} (a))for $E \lesssim 0.2E_0$. Moreover, $F_s(q,t)$ (Eqn.~\eqref{fsqt}) decays to zero at long times even at packing fraction as high as $\phi = 0.85$ ( Fig.~\ref{Fig2} (b)). The time dependence of $F_s(q,t)$ is well fit by a single stretched exponential function $F_s(q,t) = \tau_0\exp[-(t/\tau)^{\beta}]$ at all values of $\phi$  (Fig.~\ref{Fig2} (b) black solid lines). The tissue behaves like a liquid even at high packing fractions (see Fig.~\ref{Fig2} (a)). For reference, we note that the critical jamming packing fraction in mono-disperse hard sphere in 3D is $\phi_J \sim 0.64$~\cite{Jamming3D}, which implies there is absence of jamming transition in the soft tissue with $E=0.2E_0$ ($E_0 =0.001 MPa$). The relaxation time $\tau_\alpha$ (Eqn.~\eqref{tauAlpha}) increases only modestly as $\phi$ increases and is well described by the Arrhenius law (see Fig.~\ref{Fig2} (c)). A similar result is found for a 2D non-confluent tissue (see \textcolor{black}{Fig.~S11 (b) and Section IX} in SI for details).



\begin{figure}[!htpb]
\begin{center}
\includegraphics[width=0.98\columnwidth]{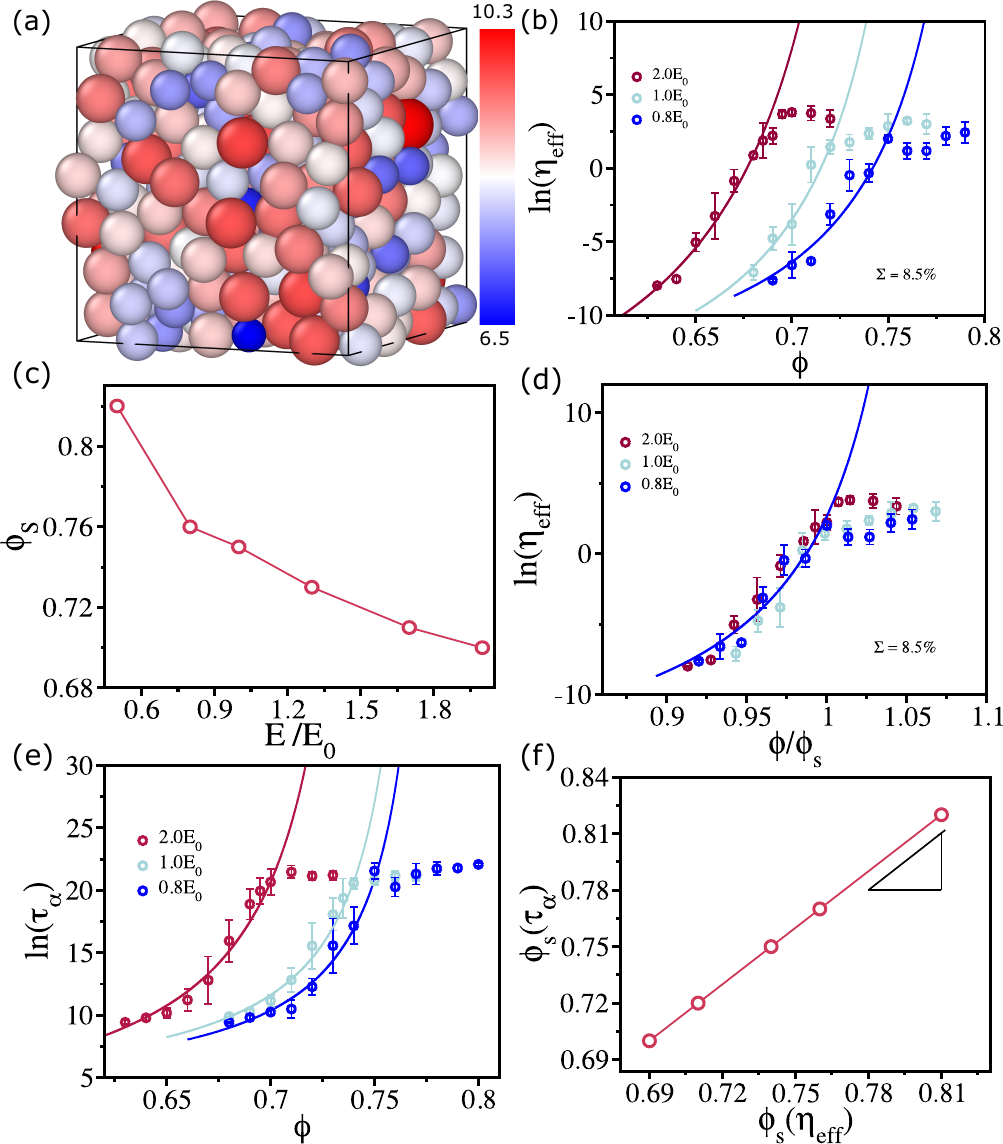}
\caption{\textbf{Viscosity at intermediate $E$:} (a) Snapshot of a 3D tissue at $\phi = 0.76$, $E= E_0$ and $\Sigma = 8.5\%$. Cells are colored according to their radii in $\mu m$ unit (shown in the color bar). (b) Effective viscosity $\eta_{\text{eff}}$ of the tissue as a function of  $\phi$ in a log-linear scale for $E = 2.0E_0, 1.0E_0$ and $0.8E_0$. The solid lines are the fits to the VFT equation (Eqn.~\eqref{VFT}). (c) Saturation packing fraction $\phi_S$ as a function of $E$. (d) The logarithm of effective viscosity $\eta_{\text{eff}}$ as a function of $\phi$ scaled by  $\phi_S$. (e) Relaxation time $\tau_\alpha$, calculated using Eqn.~\eqref{tauAlpha}, as a function of $\phi$ for $E_i = 2.0E_0, 1.0E_0$ and $0.8E_0$.  VFT fits are given by the solid line (Eqn.~\eqref{VFT} for $\tau_\alpha$). (f) $\phi_S$ calculated using $\tau_\alpha$ and $\eta_{\text{eff}}$ are linearly related. }
\label{Fig1}
\end{center}
\vskip -0.5cm
\end{figure}

\textbf{Viscosity Saturation (VS) regime:} 
Next, we simulated a system with $E = E_0 = 10^{-3} MPa$ in order to assess if a small value of $\Sigma \sim 8.5\%$  (see image of the tissue in Fig.~\ref{Fig1} (a)) captures the saturation of $\eta_{\text{eff}}$~\cite{PETRIDOU20211914} at high $\phi$.
We calculated the effective viscosity ($\eta_{\text{eff}}$)  using the Green-Kubo like relation~\cite{Hansen-McDonaldbook}.   The numerical procedure used previously~\cite{Das2023}. We define $\eta_{\text{eff}}$ as, 
\begin{equation}
    \Bar{\eta}  = \int_0^{\infty}dt \sum_{(\mu\nu)}\left\langle P_{\mu\nu}(t)P_{\mu\nu}(0) \right\rangle.
    \label{greenKubob}
\end{equation}
The stress tensor $P_{\mu\nu}(t)$ is,
\begin{equation}
P_{\mu\nu}(t) = \frac{1}{V}\left(  \sum_{i=1}^N\sum_{j>i}^N \vec{r}_{ij,\mu}\vec{f}_{ij,\nu}\right),
\end{equation}
where $\mu,\nu \in (x,y,z)$ are the Cartesian components of the vector,  $\vec{r}_{ij} = \vec{r}_i - \vec{r}_j $, $\vec{f}_{ij}$ is the force between $i^{th}$ and $j^{th}$ cells and $V$ is the volume of the simulation box.  Note that $\Bar\eta$ should be viewed as a proxy for shear viscosity because it does not contain the kinetic term and the factor $\tfrac{V}{k_BT}$ is not included in  Eqn.~\eqref{greenKubob} because temperature is not a relevant variable in the highly over-damped model considered here.
The effective viscosity of the 3D tissue as a function of $\phi$ (see Fig.~\ref{Fig1} (b)) has the same trend as in 2D: $\eta_{\text{eff}}$ increases rapidly following Vogel-Fulcher Tamman (VFT)  relation~\cite{Tammann,Fulcher}, which is given by, 
\begin{equation}
    \eta_{\text{eff}} =\eta_0\exp\left[ \frac{D}{\phi_0/\phi - 1}\right],
    \label{VFT}
\end{equation}
where $D$ is a material dependent parameter, and $\phi_0$ is the cell volume fraction at which $\eta_{\text{eff}}$ is expected to diverge. 
The VFT relation holds till $\phi \sim \phi_S$. When $\phi$ exceeds $\phi_S$, we find that $\eta_{\text{eff}}$ saturates, as observed in experiments~\cite{PETRIDOU20211914}. 

We next varied the cell softness to estimate the range of $E$ values over which the VS behavior holds at the fixed value of $\Sigma$ ($= 8.5\%$). The dependence of $\eta_{\text{eff}}(\phi)$ on $\phi$ at two other values of $E$, $0.8E_0$ and $2.0E_0$ also show the VS behavior (see Fig.~\ref{Fig1} (b)). However, the value of $\phi_S$ varies substantially as $E$ is changed (see Fig.~\ref{Fig1} (c)). Softer tissues have larger $\phi_S$ values, which is a reflection of the  ability of cells to interpenetrate each other to a greater extent than cells with larger $E$ values. Importantly, the effective viscosity $\eta_{\text{eff}}$ as a function of $\phi$ for different values of $E$ fall on a single curve when $\phi$ is scaled by the saturation packing fraction $\phi_S$ (see Fig.~\ref{Fig1} (d)). 

\textbf{Estimating $\phi_S$ from relaxation times:} Because the calculation of $\eta_{\text{eff}}$ is computationally intensive, we explored the possibility of estimating $\phi_S$  using the  relaxation time ($\tau_\alpha$) as a proxy. 
The dependence of $\ln(\tau_\alpha)$ as a function of $\phi$ at the three $E$ values are qualitatively similar to $\ln(\eta_{\text{eff}})$ versus $\phi$ (compare Fig.~\ref{Fig1} (b) and Fig.~\ref{Fig1} (e)). The relaxation time follows the VFT law (Eqn.~\eqref{VFT}) till $\phi \simeq \phi_S$ and also saturates upon further increase in $\phi$ (Fig.~\ref{Fig1} (e)). The values of $\phi_S$  (Fig.~\ref{Fig1} (e))  nearly coincide with the ones calculated from the data in Fig.~\ref{Fig1} (b). Fig.~\ref{Fig1} (f) shows that $\phi_S$ obtained using $\eta_{\text{eff}}$ and $\tau_\alpha$ are linearly related with a slope that is near unity. Furthermore, we find that $\eta_{\text{eff}} \propto \tau_\alpha$ (see \textcolor{black}{Fig.~S2 and Section II in} SI for details). Consequently, the use of $\tau_\alpha$ as a proxy in the calculation of $\phi_S$ is justified. The results in Figures ~\ref{Fig1} (b) show that $\eta_{\text{eff}}$ (see also Fig.~\ref{Fig1} (e)) as well as $\phi_S$ (see Fig.~\ref{Fig1} (c)) change substantially as the cell softness changes. 

\begin{figure}[!htpb]
\begin{center}
\includegraphics[width=0.98\columnwidth]{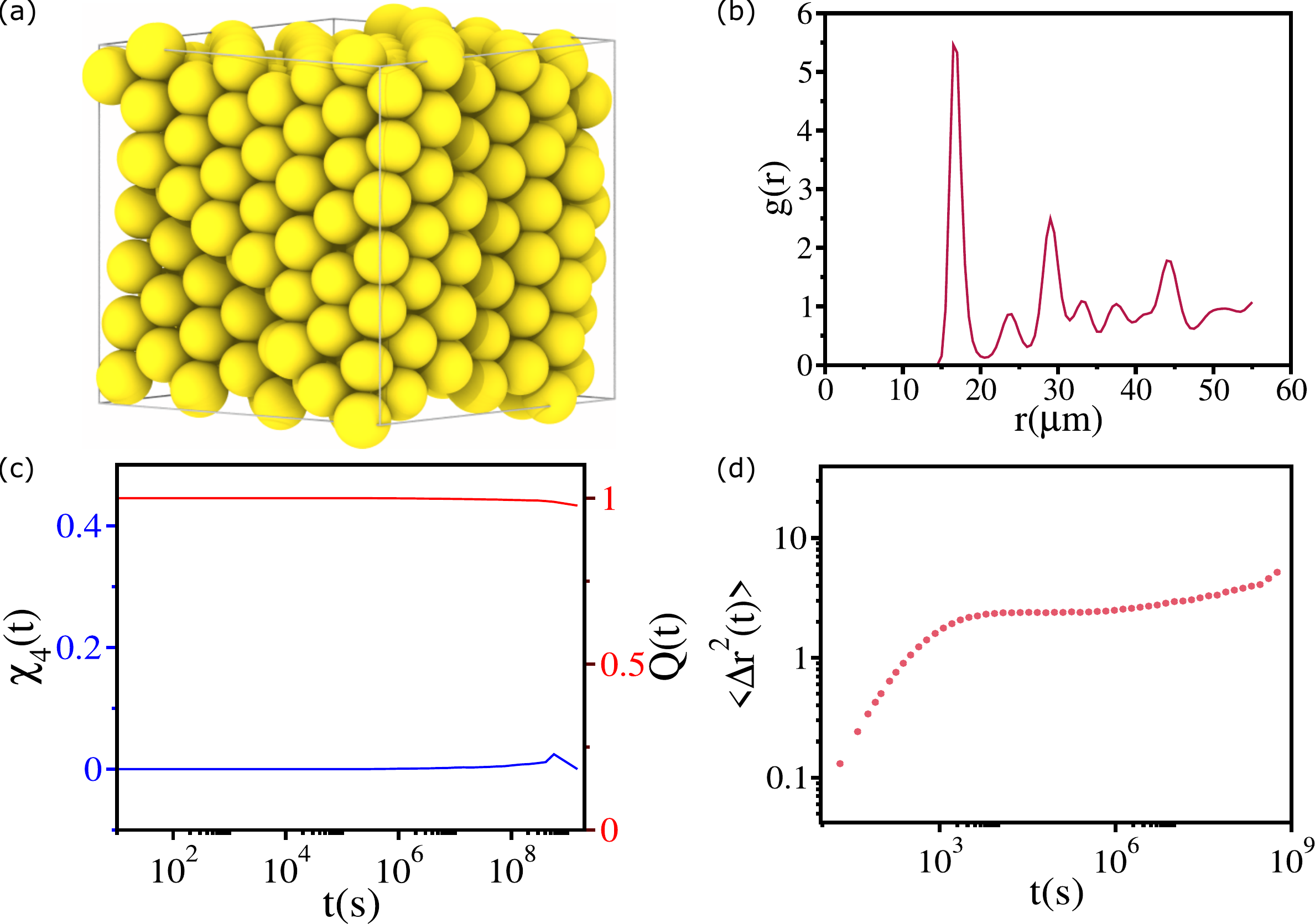}
\caption{\textbf{Crystallization:} (a) Snapshot of the 3D tissue for $\Sigma \simeq 4\%$ at $\phi = 0.77$ for $E = E_0$. (b) Pair correlation function $g(r)$ as a function of $r$ for the parameters in (a).  (c) $Q(t)$  and $\chi_4(t)$ (Eqn.~\eqref{chi4}) as a function of $t$. (d) Mean square displacement, MSD (Eqn.~\eqref{msd}), $\left\langle \Delta r^2(t) \right\rangle$ as a function of $t$.}
\label{Fig3}
\end{center}
\end{figure}
\textbf{Onset of Crystallization:}
The liquid and viscosity saturation (VS) regimes are observed if $\Sigma$ is  at or above $8.5\%$. 
To investigate the impact of lower values of $\Sigma$ on the tissue morphology, we conducted simulations at $E = E_0$ $(10^{-3} MPa)$ with $\Sigma \simeq 4\%$. The tissue crystallizes, a significant shift from the more disordered states associated with higher values of $\Sigma$. This transition is visually apparent in the images (Fig.~\ref{Fig3} (a)) and is further confirmed by the well-defined peaks in  $g(r)$ (Fig.~\ref{Fig3} (b)), indicative of a crystalline arrangement. In the crystalline state, the cells are localized near their equilibrium positions. To confirm this, we calculated the overlap function $Q(t)$ defined as $Q(t) = \left\langle\frac{1}{N}\sum_{i =1}^N w(|\Vec{r_i}(t) -\Vec{r_i}(0)|\right\rangle$, where $w(x) = 1$ if $x\leq 0.3 \left\langle\sigma\right\rangle$ \textcolor{black}{($\left\langle\sigma\right\rangle$ is the average cell diameter)} and $w(x) =0$ otherwise; $\left\langle ...\right\rangle$ is the ensemble average and average over different time origins. $Q(t)$ does not decay and remains at the value $\sim 1$ (Fig.~\ref{Fig3} (c). Furthermore, the fluctuation in $Q(t)$,
\begin{equation}
    \chi_4(t) = \frac{1}{N}\left( \left\langle Q^2(t)\right\rangle - \left\langle Q(t)\right\rangle^2 \right)
    \label{chi4}
\end{equation}
is independent of time and remain at the value $\sim 0$, which shows that the system is in a pure state. This is further reflected in the plot of mean squared displacement (MSD),
\begin{equation}
    \left\langle \Delta r^2(t) \right\rangle = \frac{1}{N}\sum_{i=1}^N \left\langle\left[(\Vec{r_i}(t) -\Vec{r_i}(0))^2\right] \right \rangle
    \label{msd}
\end{equation}
as a function of $t$ (Fig.~\ref{Fig3} (d). After an initial rearrangement, the cells do not move from their positions.

\begin{figure}[!htpb]
\begin{center}
\includegraphics[width=0.98\columnwidth]{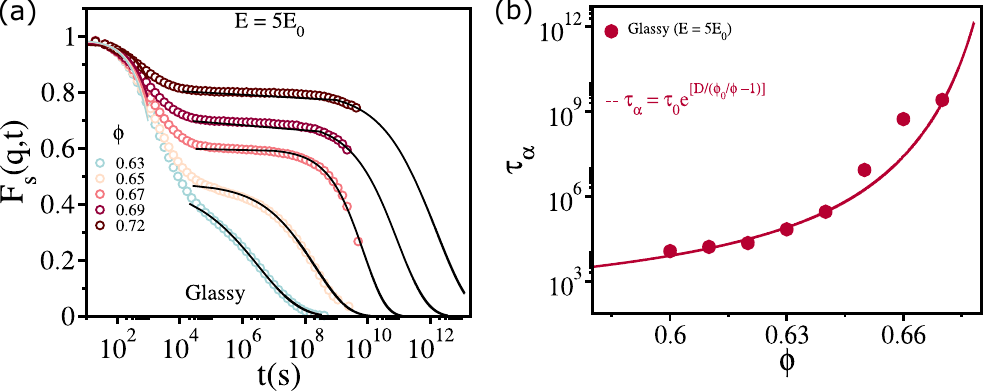}\caption{\textbf{Glassy states:} (a) Self-intermediate scattering function, $F_s(q,t)$, as a function of $t$ for $0.63 \leq \phi \leq 0.72$ for $E = 5E_0$ \textcolor{black}{($E_0\equiv 10^{-3}$MPa)}. The solid lines in color at short times are exponential fits $F_s(q,t) = \tau_0\exp[-t/\tau_s]$. The black solid lines are stretched exponential fit to $F_s(q,t) = \tau_0\exp[-(x/\tau_l)^{\beta}]$. The values of the stretching exponent, $\beta$ values are 0.34, 0.45, 0.67, 0.50, and 0.44 for $\phi$ values of 0.63, 0.65, 0.67, 0.69, and 0.72, respectively. In (a) and (b) $\Sigma = 8.5\%$. (c) $\tau_\alpha$ as a function of $\phi$ for $E=5E_0$. The solid line is the VFT fit (Eqn.~\eqref{VFT}). The fitting parameters are $\phi_0 \sim 0.70$ and $D\sim 0.78$. }    
\label{Fig2G}
\end{center}
\vskip -0.3cm
\end{figure}
\textbf{Glassy states:}
We found that when cells are extremely soft the tissue behaves like a fluid.
In the opposite limit ($E/E_0 \gg 1 $) (\textcolor{black}{$E_0 =0.001$ MPa}), the cells are very stiff and hence are not easily deformed. We expect that the system should exhibit characteristics of a hard sphere glass. Therefore, with increasing packing fraction, the effective free volume available for each cell should decrease until a critical value when the packing fraction can not be increased further. The system would be jammed, exhibiting slow dynamics. In this case, the viscosity (or relaxation time) should diverge at a high packing fraction, which is one of the characteristics of fragile glasses~\cite{ANGELL199113}, \textcolor{black}{ in which the viscosity or relaxation time as a function of $\phi$ follows the super-Arrhenius behavior (see Eqn.~\eqref{VFT}). Furthermore, the viscosity/relaxation time could diverges at a high packing fraction. }

In accord with the expectations outlined above, we find that the decay of $F_s(q,t)$ is slow, with a discernible plateau (a signature of caging-- where particles are trapped inside cages formed by their neighbors for a finite amount of time, which increases with increasing with compression) at long times, as $\phi$ is increased (Fig.~\ref{Fig2G} (a)). The scattering function decays in two distinct steps. Initially, there is a rapid exponential decay, $F_s(q,t) \sim \tau_0\exp(-\tfrac{t}{\tau_s})$ for $t \ll \tau_\alpha$, followed by a prolonged phase with stretched exponential decay, given by $F_s(q,t) \sim \tau_0\exp(-\tfrac{t}{\tau_l})^{\beta}$ for $t \sim \tau_\alpha$  (Fig.\ref{Fig2G} (a)). The microscopic relaxation time is $\tau_0$, while $\tau_s$ and $\tau_l$ are the short and long relaxation times, respectively. The values of the stretching exponent $\beta$ are 0.34, 0.45, 0.67, 0.50, and 0.44 for $\phi = 0.63, 0.65, 0.67, 0.69$, and $0.72$, respectively. Interestingly, $\tau_\alpha$ as a function of $\phi$ follows the VFT law (Eqn.~\eqref{VFT}).
The dependence of $\tau_l$ on $\phi$ also follows the VFT law (see Fig. S4 (a)), whereas $\tau_s$ as a function of $\phi$ is best described by the Arrhenius law, $\tau_s = \tau_0\exp(A\phi)$ (see Fig. S4 (a)).
 We surmise that at high values of $\phi$, the dynamics of the cells are best described as fragile glasses, where the viscosity ($\eta$) or the relaxation time $\tau_\alpha$ as a function of $\phi$ follows the super-Arrhenius behavior described by VFT law~\eqref{VFT} (Fig.~\ref{Fig2G} (b)).

\textbf{Effect of size polydispersity on tissue dynamics:}
\begin{figure}[!htpb]
\begin{center}
\includegraphics[width=0.98\columnwidth]{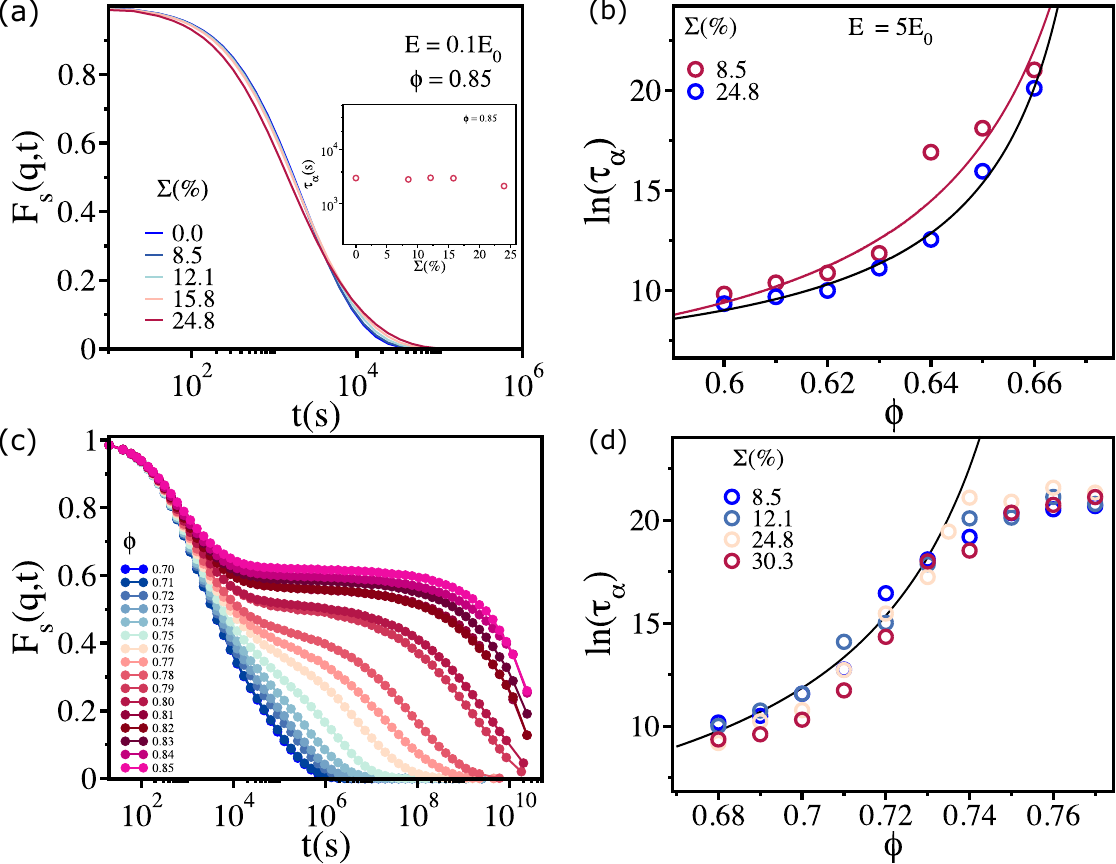}\caption{\textbf{Effect of polydispersity:} (a) $F_s(q,t)$ as a function of $t$ for $0.0 \leq \Sigma \leq 24.8$ and $E = 0.1E_0$ (liquid state) at the highest packing fraction $\phi =0.85$.  Inset shows $\tau_\alpha$ as a function of $\Sigma$ for $\phi = 0.85$. (b) Logarithm of the relaxation time $\tau_\alpha$ as a function of $\phi$ for $\Sigma = 8.5$ and $\Sigma = 24.8$ at $E = 5E_0$ (glassy state). The solid lines are VFT fits (Eqn.~\eqref{VFT}). $\phi_{0} \simeq 0.69$ and $\simeq 0.68$ for $\Sigma = 8.5$ and $\Sigma = 24.8$ respectively. 
(c) $F_s(q,t)$ as a function of $t$ for $0.70 \leq \phi \leq 0.85$ for $E = 0.5E_0$ \textcolor{black}{with $\Sigma = 24.8\%$}. (d) The logarithm of the relaxation time $\tau_\alpha$ as a function of $\phi$ for $8.5\% \leq \Sigma \leq 30.3\%$. The solid line is the VFT fit (Eqn.~\eqref{VFT}) to the data for $\phi \leq \phi_S$. } 
\label{Fig2New}
\end{center}
\vskip -0.3cm
\end{figure}

How are the dynamics affected when $\Sigma$ is varied while maintaining cell softness $E$ fixed at a particular value? Not expectedly, the answer to this question depends on the cell softness. When the tissue behaves like a liquid (cells are soft), the $\Sigma$ has a minimal influence on the dynamics. The self-intermediate scattering function $F_s(q,t)$ does not change significantly upon a large variation in $\Sigma$ even at high packing fraction $\phi = 0.85$ (Fig.~\ref{Fig2New} (a)). This is reflected in $\tau_\alpha$ as a function of $\Sigma$ (Fig.~\ref{Fig2New} (a) inset). We tested a similar effect on the glassy region. Rather than directly measuring the effective viscosity $\eta_{\text{eff}}$, we computed the relaxation time $\tau_\alpha$ to measure the effect of $\Sigma$ on the tissue dynamics. We find that qualitative behavior does not change; however, $\phi_{0}$ changes with $\Sigma$ (Fig.~\ref{Fig2New} (b).

Next, we simulated the tissue at $E =0.5E_0$ and varied $\Sigma$ in the range $8.5\% \leq \Sigma \leq 30.3\%$. 
We find that $F_s(q,t)$ shows systematic slowing down for $\phi \leq \phi_S$, and beyond that, essentially all the curves fall on top of each other (Fig.~\ref{Fig2New} (c)). This reflects the saturation in the viscosity, established using the dependence of $\eta_{\text{eff}}$ on $\phi$. 
Interestingly, we find that once $\Sigma$ exceeds a threshold of approximately $8.5\%$, the qualitative nature of the VS behavior remains unaffected. The $\phi_S$ value does not change with changing $\Sigma$ (Fig.~\ref{Fig2New} (d)). This behavior is consistent across different values of the elastic modulus $E$ where the tissue exhibited VS behavior.

\textbf{Morphology changes as a function of polydispersity and softness:}
\label{strutureLiquids}
\begin{figure*}[ht!]
\centering
\includegraphics[width = 0.98\textwidth]{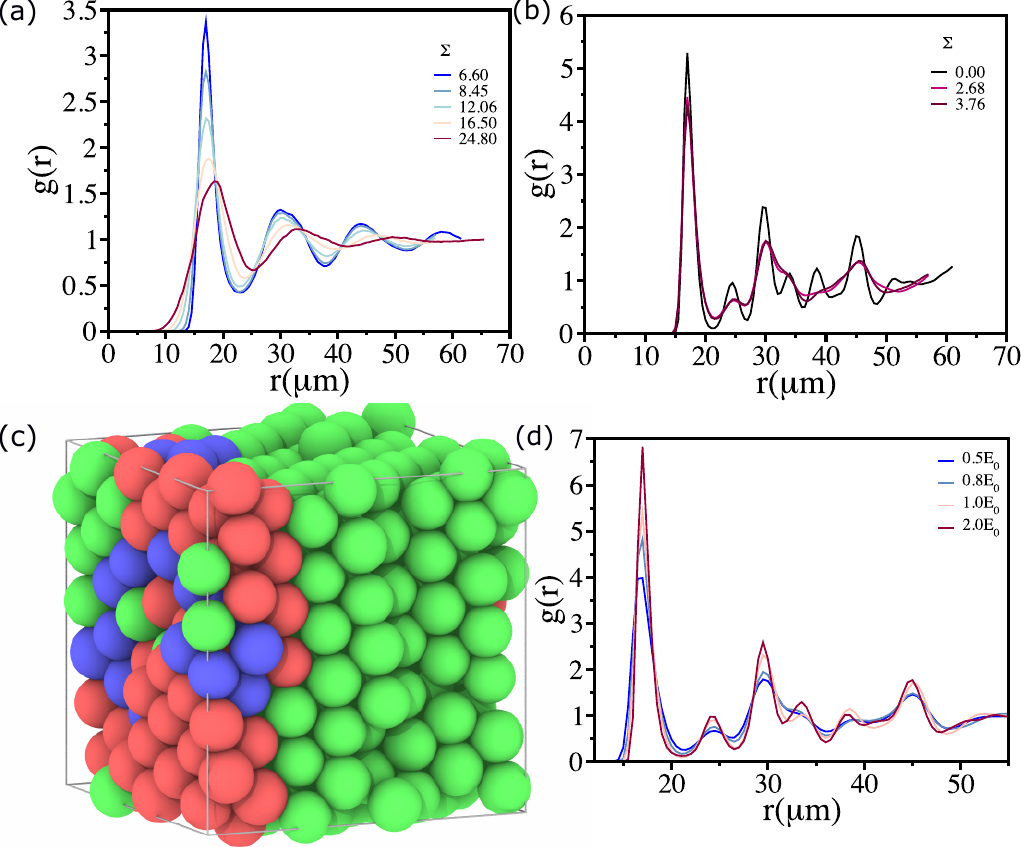}
\caption{\textbf{Structure of the 3D tissue:} Pair correlation function $g(r)$  as a function of $r$ at $\phi =0.70$ for $6.6 \leq \Sigma \leq 24.8$ with $E = E_0$. (b) Same as (a) except the range is $0.0 \leq \Sigma \leq 3.76$. (c) Snapshot for $\Sigma = 2.68$ and $E_i = 1.0E_0$. The green, blue, and red color corresponds to FCC, BCC, and HCP crystals, respectively using a-CNA analysis (see SI section XII for details). (d) $g(r)$ for $0.5E_0 \leq E_i \leq 2.0E_0$ with $\Sigma = 0.0$.}
\label{structure}
\end{figure*}
The non-confluent 3D tissue exhibits similar amorphous structures in the liquid, glassy, and viscosity saturation regimes, as reflected in the pair correlation function $g(r)$ (see Fig. S1 in SI). There is no long-range order. However, we anticipate that at a fixed $E$, polydispersity could change the structure of the tissue. In order to investigate the effect of $\Sigma$ on the tissue architecture, we calculated $g(r)$ at $\phi = 0.70$ for $6.6 \leq \Sigma \leq 24.8$ with a fixed $E = E_0$ (VS regime) (Fig.~\ref{structure} (a)). Interestingly, we find that even though the structures are amorphous, there are quantitative differences. As $\Sigma$ decreases, the magnitude of the first peak in the $g(r)$ increases systematically. Furthermore, the peaks become sharper as $\Sigma$ decreases. This also holds good for the second, third, and fourth peaks. Note that the fourth peak disappears at $\Sigma \geq 12\%$. This clearly indicates that the tissue approaches a more ordered structure as $\Sigma$ decreases. A similar behavior is found in the glassy and liquid states.

Pair functions ($g(r)$) for $\Sigma = 3.76,2.68$ and $0.0\%$ at $\phi =0.70$ with $E = E_0$ show that there is a significant change in the morphology even in the crystalline states (Fig.~\ref{structure} (b)). As $\Sigma$ decreases, the tissue acquires a perfect crystalline order: the peaks in the $g(r)$ are more prominent (the black line in Fig.~\ref{structure} (b)). An image for $\Sigma = 2.68\%$ (Fig.~\ref{structure} (c)) shows that the crystalline state is a mixture of different crystal types, such as FCC, BCC, HCP, categorized using Common Neighbor analysis method~\cite{Honeycutt1987,Stukowski2012} (see SI section VIII for details) in OVITO~\cite{ovito}. It is not a single crystal. Therefore, for a fixed value of cell softness, the polydispersity alone alters the morphology of the tissue dramatically. 

The next question is, how does morphology change with the mechanical properties of the cells, such as cell softness? To answer this, we consider the simplest case with $\Sigma = 0\%$. For a hard sphere liquid, one would expect a perfect FCC crystal for $\Sigma =0\%$.
For $\Sigma =0\%$, we computed $g(r)$ for $0.5E_0 \leq E \leq 2.0E_0$ at $\phi = 0.70$ (Fig.~\ref{structure} (d)). Interestingly, as $E$ increases, the magnitude of the first peak in the $g(r)$ increases systematically. Furthermore, the width of the peaks become narrower with increasing $E$. This clearly indicates that as the cells become rigid, the tissue approaches a near perfect ordered structure (FCC crystal see Fig. S10 (a)). 
Therefore, even for a monodisperse tissue, the morphology strongly depends on the cell rigidity. 

\begin{figure}[!htpb]
\begin{center}
\includegraphics[width = 0.95\columnwidth]{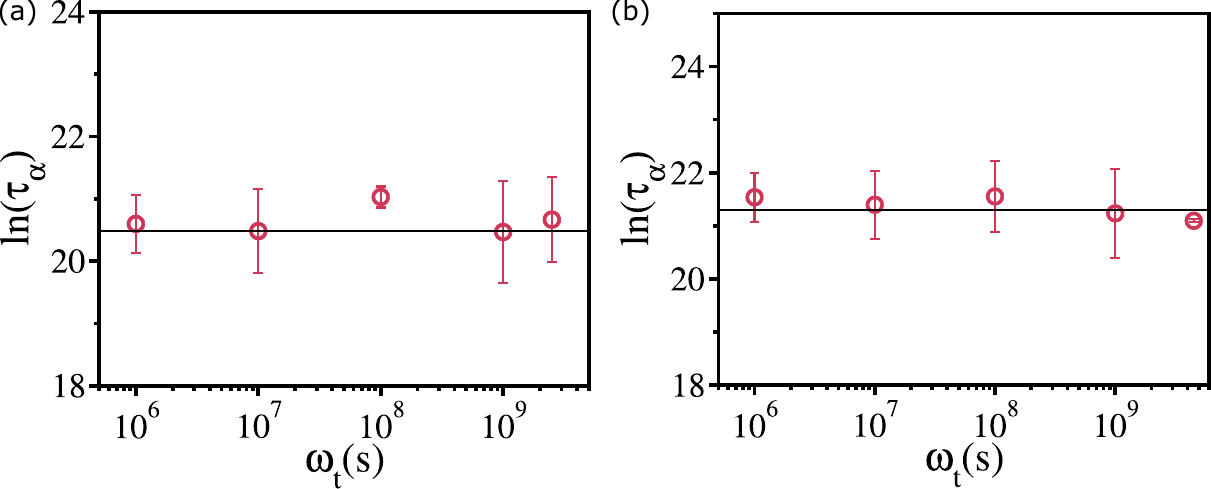}
\caption{\textit{Absence of aging in the 3D tissue:} (a) Relaxation time $\tau_\alpha$ as a function of the waiting time, $\omega_t$, at $\phi = 0.75$ for $E = 0.8E_0$ and $\Sigma = 8.5\%$. The solid line, which is a guide to the eye,  shows that $\tau_\alpha$ is independent of $\omega_t$. (b) Same as (a), except it is for $\phi = 0.66$ with $E = 5E_0$ and $\Sigma = 8.5\%$. (a) Corresponds to the VS regime and (b) is for the glassy regime.}
\label{aging}
\end{center}
\end{figure}
\begin{figure}[!htpb]
\begin{center}
\includegraphics[width = 0.85\columnwidth]{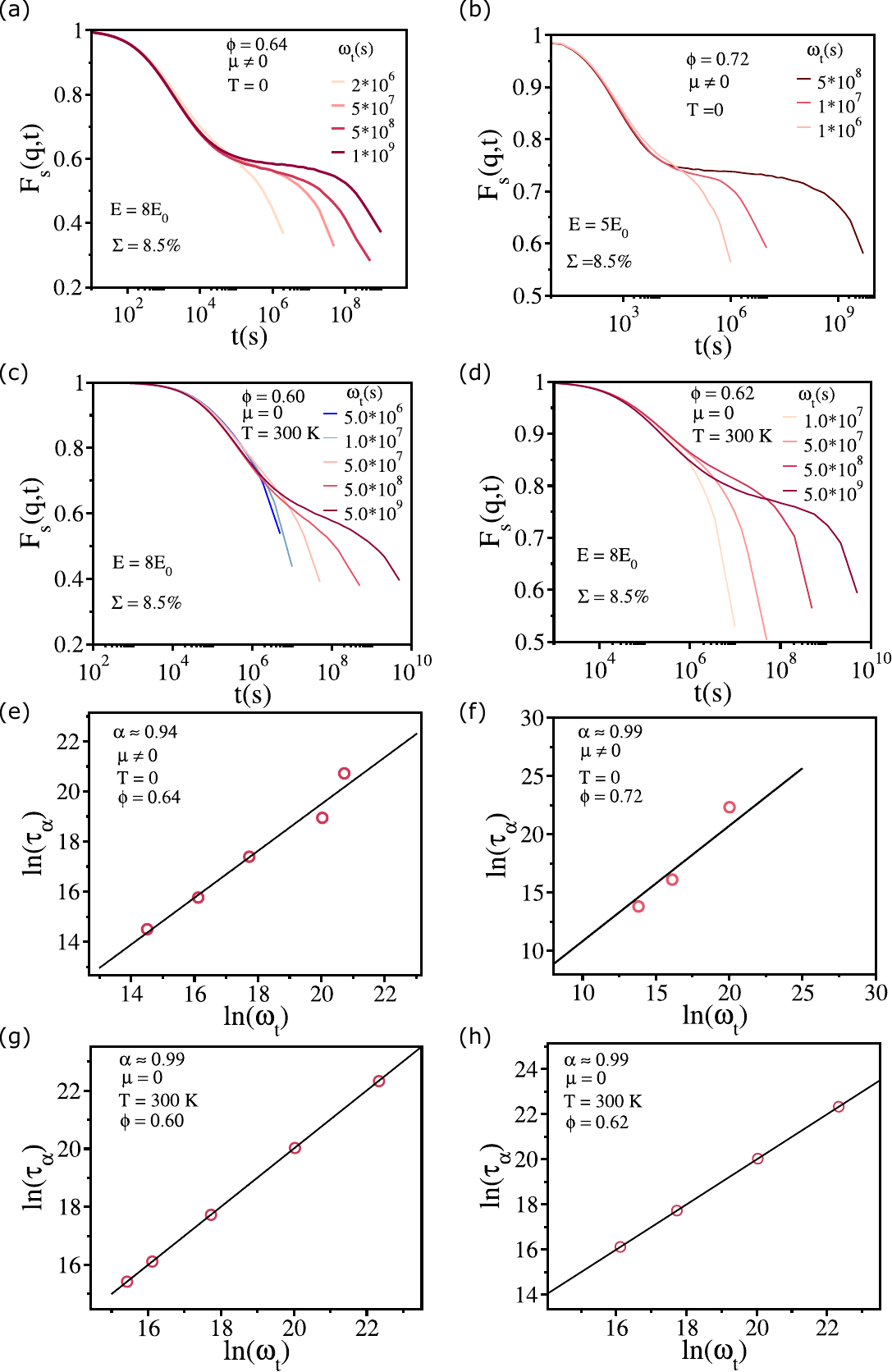}
\caption{\textit{Aging in the 3D tissue:} (a) Self-intermediate scattering function $F_s(q,t)$ as a function of the waiting time, $\omega_t$, at $\phi = 0.64$ for $E = 8E_0$ and $\Sigma = 8.5\%$.  (b) Same as (a), except it is for $\phi = 0.72$ with $E = 5E_0$ and $\Sigma = 8.5\%$. (a) and (b) are for systems with self-propulsion. (c) Same as (b) without any self-propulsion at temperature $T=300 K$ and at $\phi =0.60$. (d) Same as (c) but for $\phi = 0.62$. Logarithm of $\tau_\alpha$ as a function of logarithm $\omega_t$ for the data in a (e),  for the data in b (f), for the data in c (g) and for the data in d (h).}
\label{agingGlass}
\end{center}
\end{figure}

\section{Aging in 3D tissues}
To assess if aging effects are prevalent in the tissue dynamics in the VS regime and glass-like states, we calculated the relaxation time $\tau_\alpha$ for different waiting times, $\omega_t$. In the VS regime ($\phi = 0.75$, $E = 0.8E_0$ and $\Sigma = 8.5\%$), the relaxation time $\tau_\alpha$ is independent of $\omega_t$ (Fig.~\ref{aging} (a)). Even upon varying  $\omega_t$ by over three orders of magnitude, the relaxation time $\tau_\alpha$ is unaffected.   Strikingly, there is no aging in even in the glassy regime ($\phi = 0.66$, $E = 5E_0$ and $\Sigma = 8.5\%$) also (Fig.~\ref{aging} (b)) at least in the simulated range of $\omega_t$, as long $E$ is not large. 

The absence of aging, shown in Fig.~\ref{aging}, is surprising because it is known that glassy systems age~\cite{Kob97PRL,Amir12PNAS,Bouchad92JPhysI,Struik77PolySciEngg},  which is reflected in the increase in the relaxation time as the waiting time is increased.   Although aging is absent in the range of $E$ probed in (Fig.~\ref{aging} (b)), we expect that as the cell stiffness, $E$,  increases it would acquire the characteristics of hard sphere glasses, leading to aging behavior by  falling out of equilibrium.   In accord with this expectation, there is evidence of aging (Fig.~\ref{agingGlass} (a) and (b)) even at low packing fraction $\phi =0.64$ (Fig.~\ref{agingGlass} (a)) in the presence of active forces (non-zero $\mu$) at sufficiently large values of $E$. 

We also calculated the dependence of $\tau$ on $\omega_t$ at a fixed $T$ at two values of $\phi$ by integrating the equations of motion in Eqn. (4) in the SI. Not unexpectedly, the systems consisting of still stiff cells ($E=8E_0$) also show signs of aging at low packing fractions (Fig.~\ref{agingGlass} (c) and Fig.~\ref{agingGlass} (d)), like their active counterparts.   It should be noted that aging effects are more pronounced (the duration of the plateau in $F_s(q,t)$ is greater)  in the presence of active forces ($\mu \ne 0$) compared to thermally activated systems. (Fig.~\ref{agingGlass}).  


Following concepts in glasses~\cite{Struik77PolySciEngg}, aging effects may by analyzed using the aging exponent, $\alpha$, (denoted as $\mu$ in the glass literature),
\begin{equation}
    \alpha = \frac{d\ln \tau_{\alpha}}{d\ln t_{\omega}}.
    \label{AgingExp}
\end{equation}
The value of $\alpha$ is for data in Fig.~\ref{agingGlass} (a), (Fig.~\ref{agingGlass} (b)), (Fig.~\ref{agingGlass} (c)) and (Fig.~\ref{agingGlass} (d) $\sim 0.9$ ($\sim 1$). These values are larger than for in  most synthetic polymers but are within the theoretically expected range ($0 < \alpha \le 1$). Strikingly, in the model of tissues used here, there is a transition from ergodic behavior (absence of aging) to non-ergodic behavior (emergence of aging) by tuning the cell elasticity.



\begin{figure}[ht!]
\begin{center}
\includegraphics[width = 0.70\textwidth]{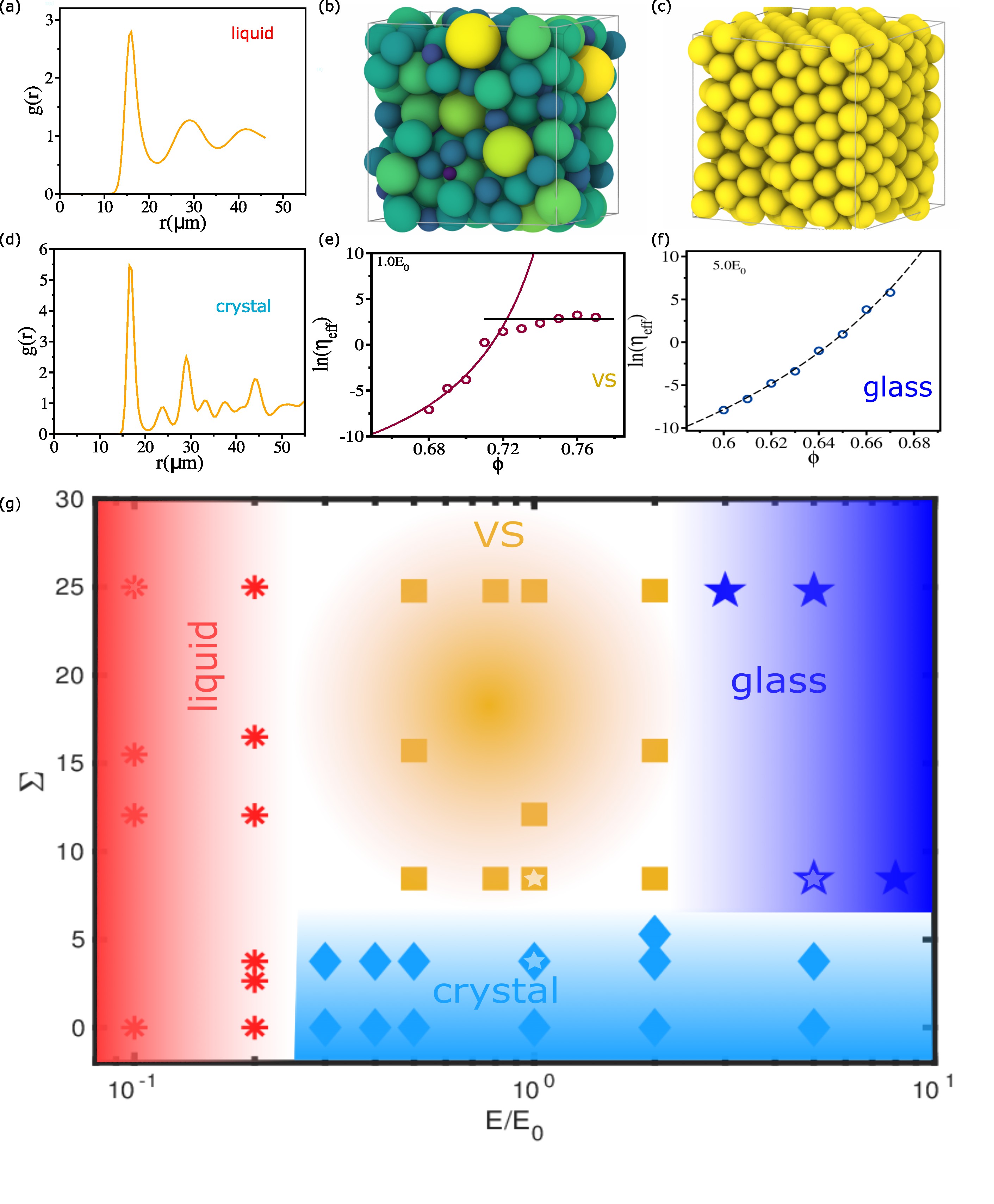}
\caption{\textbf{Phase diagram of 3D tissue:} (a) The pair correlation function $g(r)$ for a typical liquid state. (b) A snapshot of a liquid-like disordered structure. (c)  An image of the crystal structure. (d) Pair correlation function $g(r)$ for the crystal state. (e) The logarithm of effective viscosity $\ln(\eta_{\text{eff}})$ as a function of $\phi$. The solid red line is the VFT fit. The horizontal black line shows viscosity saturation. (f) The logarithm of effective viscosity $\ln(\eta_{\text{eff}})$ as a function of $\phi$. VFT fit is shown as a dashed line. (g)  Phases in the $\Sigma-E_i$ plane. The red star represents liquid states; the cyan diamond shows crystalline states; the blue corresponds to a glassy state; and the yellow square is in the viscosity saturation regime. (a) and (b) correspond to the red open star in (g), (c) and (d)  to the open cyan diamond in (g), (e) represents the open square in (g), and (f) correspond to to the open pentagram in (g).}
\label{Fig4}
\end{center}
\end{figure}
\textbf{Diagram of states:} Using both morphology and dynamics as function of $E$ and $\Sigma$, we construct a phase diagram as a function of $E$ and $\Sigma$ encompassing  liquid state, a crystalline state, a glassy state, and a viscosity saturation regime (Figure \ref{Fig4}). When the cells are soft ($E \lesssim 0.20E_0$), the tissue behaves as a liquid, characterized by structure (Fig.~\ref{Fig4} (a) and (b)) and dynamics \textcolor{black}{(Fig.~\ref{Fig2} (a), (b)}), irrespective of the value of $\Sigma$. This behavior arises from the cells being so soft that they can readily pass through one another. It is noteworthy that even in the absence of polydispersity, the monodisperse system does not crystallize under these conditions (\textcolor{black}{see Fig.~\ref{Fig2}} (c)). The crystalline state is characterized by structure (Fig.~\ref{Fig4} (c) and (d)) and dynamics (Fig.~\ref{Fig3} (c) and (d)). Even in the $\Sigma =0\%$ case, a liquid-to-crystal transition occurs near $E \sim 0.3E_0$. However, the exact value of $E_0$ where the transition happens is difficult to calculate in the current scope of study. For a finite value of $\Sigma \lesssim 6\%$ also, this transition seems to appear near $E \sim 0.3E_0$.

When $0.5E_0 \lesssim E \lesssim 2.0E_0$, a crystalline to a viscosity saturation regime (Fig.~\ref{Fig4} (e)) transition occurs near $\Sigma \sim 8.5\%$. On the other hand, when $E \gtrsim 3.0E_0$, there is transition from a crystalline state to a glassy state (Fig.~\ref{Fig4} (f))  near $\Sigma \sim 8.5\%$. A liquid to VS transition is found  for $\Sigma \gtrsim 8.5\%$ and $E \sim 0.5E_0$.  


\section{Discussion}
We explored the link between morphologies and dynamics of non-confluent  tissues in three dimensions. The morphological changes were controlled by varying the cell elasticity ($E$) and dispersion in cell sizes $\Sigma$.  The dynamics is driven by stochastic active self-propulsion forces that are uncorrelated in space and time.  Despite the simplicity of the model, a number of distinct morphological states  emerge as $\Sigma$ and $E$ are varied.   In the absence of $\mu$ (see Eqn.~\eqref{EOM}) the cells would not move, which implies an initial structure at a given $\Sigma$ and $E$ would remain forever frozen. Thus, the complex arrangement of cells at different  $\Sigma$ and $E$ values, depicted in the phase diagram, requires non-zero $\mu$ that drives the system out of equilibrium. Nevertheless,   ideas in equilibrium statistical mechanics  and linear response theories rationalize many of the findings. For instance, the dependence of the effective viscosity calculated using Green-Kubo like relation accounts for the experimentally observed saturation in $\eta_{eff}$  on volume fraction (Fig.~\ref{Fig1} (b). In addition, crystallization at low $\Sigma$ is readily understood in terms of pair correlation functions. We close this article with a few additional remarks.   

\textit{Activity driven phases:} The interplay between cell size polydispersity and cell softness produces a range of tissue morphology in presence of active forces. In what follows, we summarize the major findings.
(a) When cells are soft ($E ~\lesssim 0.2E_0$), the tissue behaves like a liquid independent of the value of $\Sigma$.  The increase in the relaxation time is best described by the Arrhenius law (Fig.~\ref{Fig2} (b). The liquid-like characteristics are maintained at high values of $\phi$ (Fig.~\ref{Fig2} (a)). This should be contrasted with simulations of two-dimensional soft colloids~\cite{Gnan2019}, which showed that the relaxation time, at a fixed temperature, increases as the area fraction increases up to a critical value and then decreases. The differences which arises because of the presence of active force, is surprising because the elasticity in the soft colloids and in the cell system are described by the Hertz potential. 

(b) When $E_i \gtrsim 0.2E_0$, a broad range of dynamical behavior emerges depending on the cell softness and size and the extent of polydispersity.  If $\Sigma$ exceeds a minimum value, the effective viscosity increases, following the VFT law, followed by a saturation of $\eta_{\text{eff}}$ as a function of $\phi$, occurs even for a relatively small value of $\Sigma$ ($\approx 8.5\%$). To observe the VS regime, a high value of $\Sigma$ is not required as long as the crystallization is avoided. Once the size dispersity exceeds $\gtrsim 8.5\%$, the tissue dynamics do not change qualitatively upon changing $\Sigma$, even though the structures are substantially different. However, the unusual viscosity behavior occurs only for an intermediate range of cell softness. 

(c) As the cell stiffness continues to increase,  the tissue exhibits  fragile glass-like behavior that is known in synthetic materials, such as ortho-terphenyl. The whole range of $\eta_{\text{eff}}$ is well fit by the VFT equation.  

(d) For $\Sigma \lesssim 6\% $, the tissue always crystallize when $E_i \gtrsim 0.2E_0.$ Interestingly, even at $\Sigma=0$, the morphology of the crystal changes with the softness. When cells are sufficiently soft, the crystal structure is a mixture of common motifs (BCC, FCC, and HCP). Strikingly, a more perfect crystal forms when the cells becomes increasingly rigid. As $E$ increases the cells become increasingly hard-sphere like, which prevents substantial overlap. In this limit, defects or packing deficiency can not propagate, resulting in the formation of ordered FCC crystal, the ground state for hard spheres (see Fig. S10 in SI). 

\textit{On the importance of cell propulsion:} It should be stressed that the rich dynamical behavior (Arrhenius dependence of $\tau_\alpha$ on $\phi$, saturation of viscosity at intermediate values of $E$ and modest values of $\Sigma$, and VFT behavior at high E) are found by driving the system by an active self-propulsion force 
(Eqn.~\eqref{EOM}). In the absences of $\mu$, the cell system is frozen at all times because the value of $\gamma$ is so large that thermal motion is suppressed. The phases (see Fig.~\ref{Fig4}) emerge as a consequence of $\mu \neq 0$. Because Eqn.~\eqref{EOM} does not satisfy FDT, the cell system is out of equilibrium but could reach a steady state. Despite the non-equilibrium nature of dynamics,  it is surprising that the relaxation time in VS regime and glassy regime do not depend on the waiting time as long as $E$ is not too large. At sufficiently large values of $E$ relatively small $\Sigma$ values, the issues exhibits glass like behavior, including aging that is reflected in the dependence of the relaxation time of the waiting time in both the driven ($\mu \ne 0$) and the passive counterpart.  

These findings are crucial for understanding the physical mechanisms underlying tissue morphogenesis and disease progression, particularly in the context of cancer, where cell rigidity significantly influences the invasive properties of tumors  and metastasis~\cite{Cross2007,Liu2020}. The metastatic cancer cells, which are often softer and exhibit greater motility than their benign counterparts~\cite{Cross2007} would exhibit more liquid-like behavior.
Furthermore, our results show that different tissues, characterized by varying degrees of cell softness and polydispersity, would likely exhibit distinct mechanical behaviors. For tissues composed of relatively homogeneous and rigid cells, the dynamics are predicted to be more glass-like with high viscosity and slow dynamics, potentially leading to jamming at high cell densities~\cite{hunter2012physics}. Conversely, tissues with softer and more heterogeneous cell populations may demonstrate fluid-like behaviors with lower viscosities, facilitating easier rearrangement and potentially \textcolor{black}{influencing processes} such as wound healing, tissue regeneration, or cancer progression~\cite{PETRIDOU20211914,fuhs2022rigid,han2020cell}.

Moreover, the saturation of viscosity at high cell densities suggests a limit to mechanical stiffening possible within tissues, beyond which increased cell packing does not further enhance tissue rigidity. This could particularly be relevant in embryogenesis process~\cite{PETRIDOU20211914}. 

\section{Materials and Methods:}
\textbf{ The model:}  We simulated a non-confluent tissue using a particle-based cell model in three dimensions (3D)~\cite{malmi2018cell, sinha2020spatially}. The 2D results are presented in the Supplementary Information (SI).
\begin{figure}[!htpb]
\begin{center}
\includegraphics[width=0.98\columnwidth]{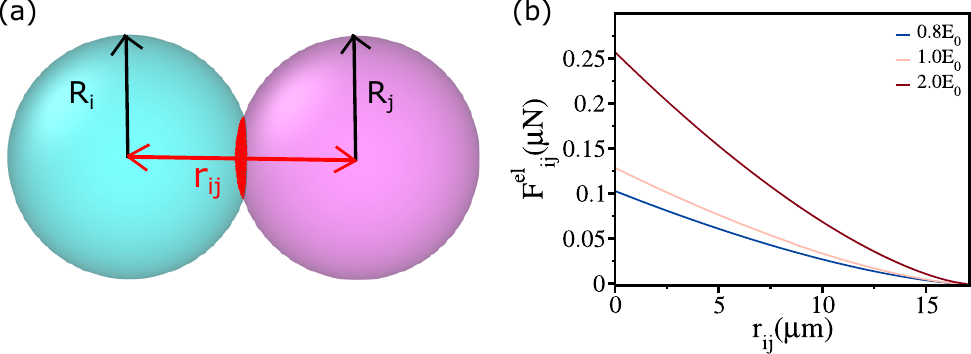}
\caption{\textbf{Hertz force:} (a) Schematic of two cells with radii $R_i = R_j = 8.5\mu m$ that overlap as indicated by the red shaded  region. (b) Hertz  force, $F^{el}_{ij}(r)$ (Eq. \ref{elastic}), as function of $r_{ij}$ for $E = 0.8E_0, 1.0E_0$ and $2.0E_0$. The value of $E_0 = 0.001 MPa$.}
\label{FigForce}
\end{center}
\end{figure}
We model cells as soft deformable spheres (Fig. \ref{FigForce}(a)) that interact via short-range forces. The Hertz elastic (repulsive) force between two cells, with radii $R_i$ and $R_j$, that are separated by a distance, $|\Vec{r_i} -\Vec{r_j}|$ is,
\begin{equation}
F_{ij}^{el} = \frac{h_{ij}^{3/2}}{\frac{3}{2}\left(\frac{1-\nu^2}{E} \right)\sqrt{\frac{1}{R_i} + \frac{1}{R_j}}},
\label{elastic}
\end{equation} 
where $h_{ij}=\text{max}[0, R_i+R_j-|\vec{r}_i-\vec{r}_j|]$.
The Hertz force acts along the unit vector $\vec{n}_{ij}$, which points from the center of the $j^{th}$ cell to the center of $i^{th}$ cell (Fig.~\ref{FigForce}a).
The total elastic force on the $i^{th}$ cell is,
$$\vec{F_i} = \sum_{j\in NN(i)}\left(F^{el}_{ij}\right)\vec{n}_{ij},$$
where $NN(i)$ is the number of \textcolor{black}{nearest} neighbor cells that are in contact with the $i^{th}$ cell. The $j^{th}$ cell is the nearest neighbor of the $i^{th}$ cell if $h_{ij} > 0$. The nearest neighbor condition ensures that the cells interpenetrate to some extent depending on the elasticity of the cell. 
For simplicity, we assume that the elastic moduli ($E$) and the Poisson ratio ($\nu$) for all the cells are identical, a condition that could be relaxed. Fig. \ref{FigForce}(b) shows $F_{ij}$ as a function of $r_{ij}$, the distance between cells $i$  and $j$. As $E$ increases, the magnitude of force increases. At a very high value of $E$, the tissue is sufficiently stiff to acquire hard sphere characteristics. 

\textbf{Dispersion in cell sizes:} We simulated samples consisting of cells of different sizes. The polydispersity (PD), $\Sigma$, is varied in the range $ 0\% \leq \Sigma \leq 30\%$,  to simulate tissues with varying morphologies.  We define PD using,
\begin{equation}
    \Sigma = \frac{\sqrt{\left\langle \sigma^2\right\rangle - \left\langle \sigma\right\rangle^2} }{\left\langle \sigma\right\rangle},
    \label{PD}
\end{equation}
where $\left\langle\sigma\right\rangle$ is the average over cell diameters. By varying $E$ and $\Sigma$, both the morphology and the tissue dynamics can be altered.

\textbf{Self-propulsion and equations of motion:} In addition to the Hertz force, we include an active force arising from the self-propulsion of individual cells  ($\mu$), which is a proxy for the intrinsically generated stresses within a cell. For simplicity, we assume that $\mu$ is independent of the cells. 
The dynamics of each cell obey the phenomenological equation,
\begin{equation}
\dot{\vec{r}}_i = \frac{\vec{F}_i}{\gamma_i} +\mu \vec{\mathcal{W}}_i(t),
\label{EOM}
\end{equation}
where $\gamma_i$ is the friction coefficient, $\gamma_i=6\pi\eta R_i$~\citep{malmi2018cell,sinha2022mechanical}, experienced by the $i^{th}$ cell, and $\mathcal{W}_i(t)$ is the active noise term. 
The statistics associated with the active uncorrelated noise, $\mathcal{W}_i(t)$, obeys  $\left\langle\mathcal{W}_i(t) \right\rangle = 0$ and $ \langle \mathcal{W}_i^{\alpha}(t)\mathcal{W}_j^{\beta}(t')\rangle=\delta(t-t')\delta_{i,j}\delta^{\alpha,\beta}$ with $\alpha,\beta \in (x,y,z)$; $\delta_{ij}$ and $\delta^{\alpha,\beta}$ are Kronecker $\delta$ functions and $\delta(t-t')$ is the Dirac $\delta$ function. 

Two pertinent comments should be made. (i) In the model, there is complete absence of dynamics with only systematic forces because the effective temperature is zero. The cell movement is generated solely by self-propulsion (Eq.~\eqref{EOM}). (ii) Notice that $\gamma_i$ and $\mu$ may be varied independently because there is no fluctuation-dissipation relation relating the two quantities. It is, in this sense, that the system is out of equilibrium at all values of $\mu$ which is taken to be $0.045 \mu m/\sqrt s$.  In the SI, we show that viscosity saturation cannot be recapitulated using equations of motion that satisfy fluctuation-dissipation relation. 


\textbf{Simulation details:} We placed N cells in a cube. To minimize finite size effects, the box is periodically replicated.  The size \textcolor{black}{($L$)} of the box is chosen such that the packing fraction is $\phi = \tfrac{\sum_{i=1}^N \pi R_i^2}{L^2}$ in 2D and $\phi = \tfrac{\sum_{i=1}^N 4\pi R_i^3}{3L^3}$ in 3D. We performed simulations by varying $\phi$ in the range $0.70 \leq \phi \leq 0.95$ in 2D and $0.60 \leq \phi \leq 0.85$ in 3D.  Because the critical jamming fraction in 3D is $\phi_J \sim 0.64$~\cite{Jamming3D}, the range of  $\phi$ explored is smaller than in 2D, where the critical jamming occurs at $\phi_J \sim 0.84$~\cite{CriticalJamming}. The parameters used in the simulations are given in Table I in the SI. The results reported in the main text are obtained with $N = 500$ in 3D. Finite size effects are discussed in the SI.

The elastic constant $E$ is reported in units of $E_0 = 0.001 MPa$. 
For each $\phi$, we performed simulations for at least ($5-10$)$\tau_\alpha$ ($\tau_{\alpha}$ is the relaxation time) before analyzing the data. For the calculation of viscosity, we perform 32 independent simulations at each $\phi$, \textcolor{black}{while 24 independent simulations are conducted for the calculation of relaxation time.} 

\noindent \textbf{Acknowledgments:} This work is supported by the National Science Foundation (grant no. PHY 2310639), the Collie-Welch Chair through the Welch Foundation (F-0019). 

\appendix

\bibliographystyle{unsrt}
\bibliography{2DCellCombined}
\end{document}


\title{Supplementary Information -- Controlling the morphologies and dynamics in three-dimensional tissues}
\author{Rajsekhar Das$^1$ }
\affiliation{Department of Chemistry, University of Texas at Austin, Austin, Texas 78712, USA}
\author{Xin Li$^1$}
\affiliation{Department of Chemistry, University of Texas at Austin, Austin, Texas 78712, USA}
\author{Sumit Sinha$^2$}
\affiliation{Department of Physics, University of Texas at Austin, Austin, Texas 78712, USA}
\author{D. Thirumalai$^{1,2}$}
\affiliation{$^1$Department of Chemistry, University of Texas at Austin, Austin, Texas 78712, USA}
\affiliation{$^2$Department of Physics, University of Texas at Austin, Austin, Texas 78712, USA}

\maketitle 


We provide a brief the description of the organization of the SI. way it is organized. The structures of the three dimensional (3D) tissue in different states are described in section (I). The relation between the viscosity ($\eta$) and the relaxation time, $\tau_\alpha$, is discussed in section (II).  In section (III), we discuss the effect of finite size on the dynamics of 3D tissue.  We describe the behavior of relaxation processes in liquid and glassy states in section (IV).  In section (V) we show that the Free volume theory explains the dynamics of 3D tissues. In section (VI), it is shown the geometric properties of cells, such as cell-cell connectivity, are related to tissue dynamics. In section (VII), we show that self-propulsion is required for the viscosity saturation.
In section (VIII), we discuss the methods used to  determine the crystalline states of a cell in the tissue. The results for 2D simulations are shown in section (IX). The parameters used in the simulations are listed in section (X).
\begin{figure}[!ht]
\begin{center}
\includegraphics[width = 0.95\columnwidth]{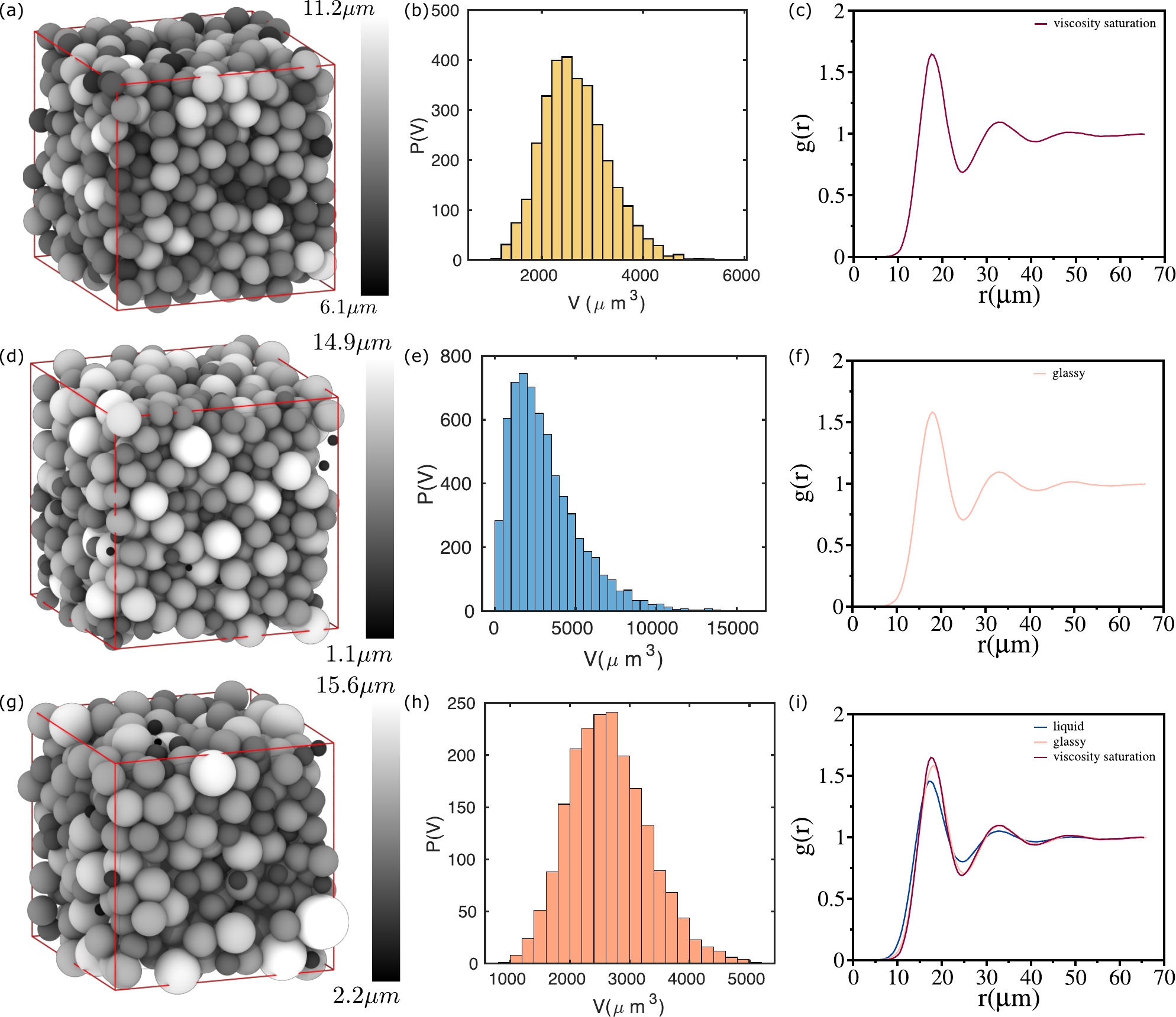}
\caption{\textit{Structures of 3D tissues:} Simulation snapshot of a 3D tissue for $E=E_0 = 0.001 MPa$, $\Sigma= 8.5\%$ (yellow shaded regime in Fig. 9 (g) in the main text), $\phi =0.70$ and $N=800$ (a), for $E = 5E_0$, $\Sigma =24\%$ (blue shaded regime in Fig. 9 (g) in the main text), $\phi =0.68$ and $N =800$ (d) and for $E = 0.2E_0$, $\Sigma = 8.5\%$ (red shaded regime in Fig. 9 (g) in the main text), $\phi=0.75$ and $N=500$ (g). Distribution of cell volume for for $E=E_0$, $\Sigma= 8.5\%$, $\phi =0.70$ and $N=800$ (b), for for $E = 5E_0$, $\Sigma =24\%$, $\phi =0.68$ and $N =800$ (e) and for  $E = 0.2E_0$, $\Sigma = 8.5\%$, $\phi=0.75$ $N =250$ (h). Pair correlation function, $g(r)$ as a function of $r$ at $\phi = 0.70$ with $\Sigma = 8.5\%$ and $N =500$ for $E =E_0$ (c), for $E = 3.0E_0$ (f) and for $E=0.2E_0$ (i) (blue curve). (i) Comparison of $g(r)$ for VS, glassy and liquid states.}
\label{struc}
\end{center}
\end{figure}
\section{Structures of 3D tissues}
Depending on the values of the cell elasticity ($E$)  and $\Sigma$, cell size dispersion, tissue dynamics exhibit the characteristics of liquids, glassy, and viscosity saturation (see Fig. 9 (g) in the main text). Even though these states are dynamically distinct, they exhibit similar disordered structures, as illustrated below.
Fig.~\ref{struc} (a) shows a snapshot of  the 3D tissue with polydispersity $\Sigma = 8.5\%$ and $E=E_0$, which corresponds to the VS regime  (yellow shaded regime in Fig. 9 (g) in the main text). The tissue exhibits an amorphous liquid like structure. The corresponding distribution of volume is shown in Fig.~\ref{struc} (b).  

We characterized the structure using the pair correlation function,
\begin{equation}
    g(r) = \frac{1}{\rho}\left\langle \frac{1}{N}\sum_i^N\sum_{j\neq i}^N \delta\left(r - |\vec{r}_i - \vec{r}_j |\right)\right\rangle
\end{equation}
where $\rho = \tfrac{N}{L^3}$ is the number density, $\delta$ is the Dirac delta function, $\vec{r}_i$ is the position of the $i^{th}$ cell, and the angular bracket $\left\langle\right\rangle$ is an ensemble average. Fig.~\ref{struc} (c) shows that there is no long-range order in the system. We also show the snapshot in the glassy regime (blue shaded regime in Fig. 9 (g) in the main text) in Fig.~\ref{struc} (d) and liquid regime ( red shaded regime in Fig. 9 (g) in the main text) in Fig.~\ref{struc} (g) with corresponding volume distribution in Fig.~\ref{struc} (e) and Fig.~\ref{struc} (h), respectively. The pair correlation function $g(r)$ for glassy and liquid regimes are shown in Fig.~\ref{struc} (f) and Fig.~\ref{struc} (i) (blue curve), respectively. However, as shown in Fig.~\ref{struc} (i), the peaks in the $g(r)$ for liquids are lower than the glass and the system showing viscosity saturation, suggesting liquids are more disordered.\\
\begin{figure}[!ht]
\centering
\includegraphics[width = 0.9\columnwidth]{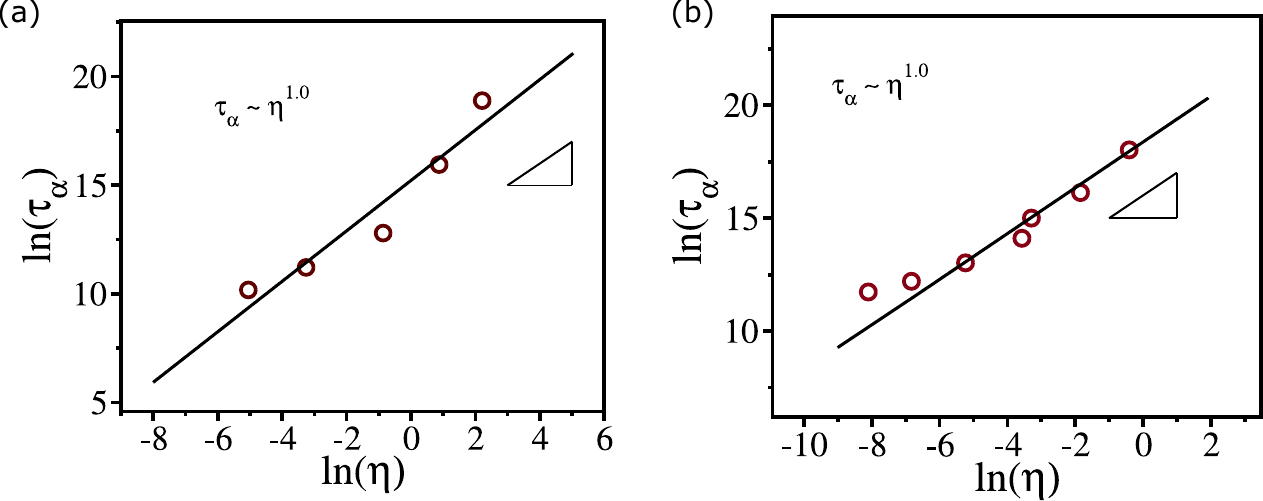}
\caption{\textit{Relation between $\eta$ and $\tau_\alpha$:} (a) Relaxation time, $\tau_\alpha$,  as a function of $\eta$ for $\phi \leq \phi_S$ for the 3D non-confluent tissue with $\Sigma =8.5\%$ and $E=E_0 $ (VS regime -- yellow shaded regime in Fig. 9 (g) in the main text). (b) Same as (a) except the results are in 2D with $\Sigma = 24\%$. Small triangles with slope 1 are a guide to the eye. }
\label{etatau}
\end{figure}
\section{Viscosity and structural relaxation time are linearly related}
Because the calculation of  $\eta$ is computationally intensive, in some instances we substituted  $\tau_\alpha$ as a proxy.   
This can only be justified if $\eta$ and $\tau_\alpha$ are linearly proportional to each other, which is not always the case in supercooled liquids~\cite{StillingerJCP}. For this reason, it is necessary to verify the relation $\eta \sim \tau_\alpha$ in the current cell model. Plots of $\tau_\alpha$ as a function of $\eta$ (for $\phi \leq \phi_S$ --  Fig. 2 in the main text) on a log-log scale (Fig.~\ref{etatau} (a) and (b)) show that $\tau_\alpha$ and $\eta$ are proportional to each other both in 3D (Fig.~\ref{etatau} (a))  and 2D (Fig.~\ref{etatau} (b)). Therefore, using $\tau_\alpha$ as a proxy for  $\eta$  is justified.

\begin{figure}[!ht]
\begin{center}
\vskip 1.cm
\includegraphics[width = 0.98\columnwidth]{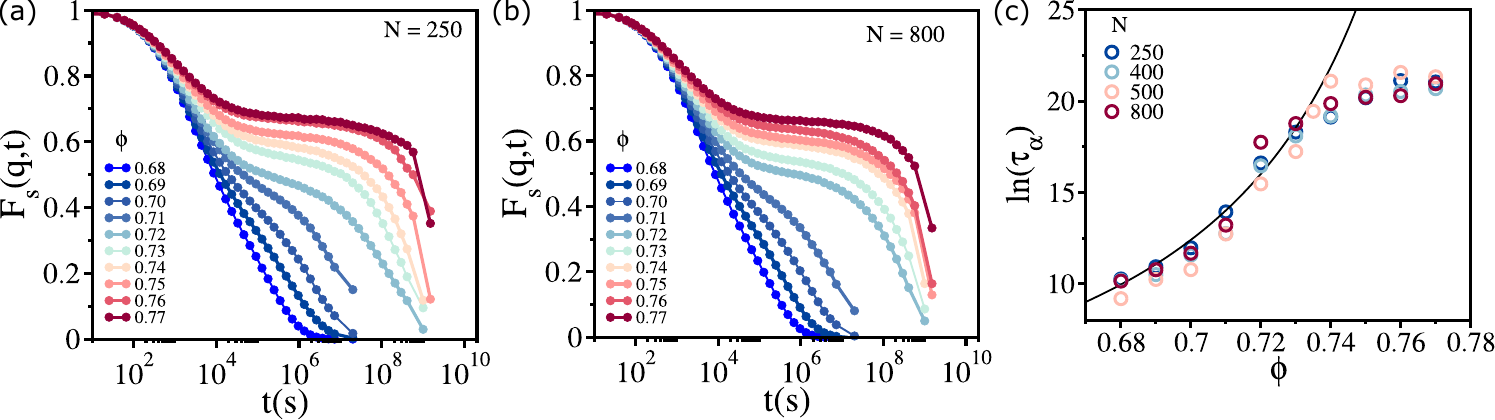}
\caption{\textit{Finite system size effects in 3D tissues:} The self-intermediate scattering function, $F_s(q,t)$ (Eqn. (1) in the main text), as a function of $t$ for $0.68 \leq \phi \leq 0.77$ with $E = E_0$ and $\Sigma = 8.5\%$.  The value of $N = 250$ (a) and $N = 800$ (b). Relaxation time $\tau_\alpha$ as a function of $\phi$ for $N = 250, 400, 500$ and $800$. The solid line is a fit to the VFT equation (see Eqn.~\ref{VFT}.)  }
\label{Finite}
\end{center}
\end{figure}
\section{Finite size effects}

In the main text, we reported the results for $N = 500$.  To assess the effects of the finite system size on tissue dynamics, particularly in the VS regime (yellow shaded regime in Fig. 9 (g) in the main text), we performed simulations using $N = 250, 400$, 500, and $800$. The variation in $N$ is not too large because the computations are numerically difficult to carry out for $N$ that is significantly greater than $800$. The results for $F_s(q,t)$ are presented in Fig.~\ref{Finite} (a) and (b) for $N = 250 $ and $800$ respectively. Fig.~\ref{Finite} (c) shows that $\tau_{\alpha}$ and $\phi_S$ ($\sim 0.74$) values are independent of $N$ in the simulated range. Because the $N$ values were varied only over a very limited range (for computational reasons), we cannot be certain that the results are truly independent of system size.

\section{ Short and long relaxation times in liquid and glassy phases }
\begin{figure}[!htpb]
\begin{center}
\vskip -0.2cm
\includegraphics[width = 0.95\columnwidth]{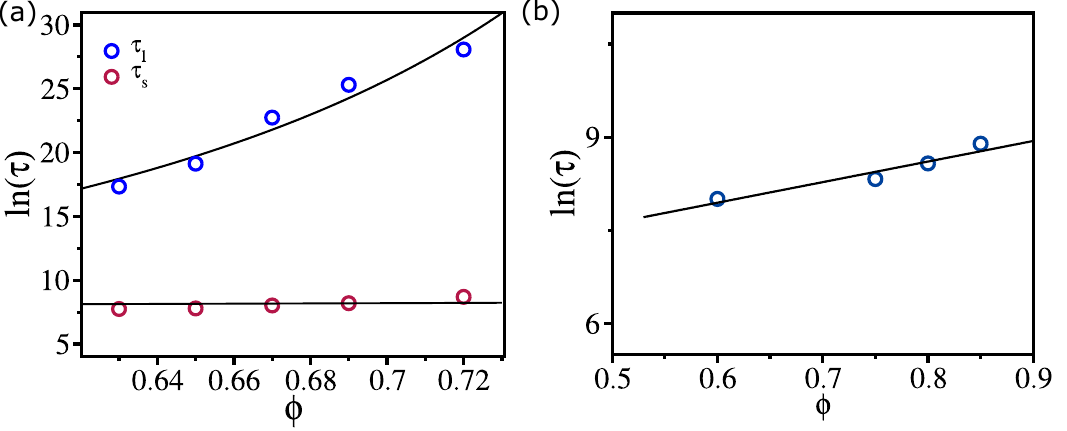}
\caption{\textit{Relaxation times as a function of $\phi$:} (a) Short ($\tau_s$) and long ($\tau_l$) relaxation times as a function of $\phi$ calculated using the data Fig. 4 (a) in the main text (glassy dynamics for $\Sigma = 8.5\%$ and $E = 5E_0$). Although $\tau_l$ is well fit by the VFT equation (Eqn.~\eqref{VFT}), $\tau_s$ follow the Arrhenius law, $\tau_s = \tau_0\exp(A\phi)$. (b) The relaxation time $\tau$, calculated from Fig. 1 (a) main text, as a function of $\phi$ in the liquid phase ($\Sigma = 8.5\%$ and $E = 0.2E_0$).  The increase in $\tau$ follows the Arrhenius law $\tau = \tau_0\exp(A\phi)$ (solid line). 
}
\label{fittedTau}
\end{center}
\end{figure}

In the liquid phase, the self-intermediate scattering function $F_s(q,t)$ decays in a single step, as shown in Fig. 1 (b) in the main text. The dependence of $F_s(q,t)$ on $t$ is fit by a single stretched exponential function (Fig. 1 (b) in the main text).
The estimated relaxation time $\tau$ follows the Arrhenius law $\tau = \tau_0\exp(A\phi)$ (Fig.~\ref{fittedTau} (b)).

In the glassy state,  $F_s(q,t)$ decays in two steps. Initially, there is a fast relaxation, and at long times, there is a slow stretched exponential decay (see Fig. 4 (a) in the main text). We calculated the short and long relaxation times $\tau_s$ and $\tau_l$ using the fitting procedure described in the main text.  Fig.~\ref{fittedTau} (a) shows that $\tau_l$ grows as $\phi$ increases, and well fit by the VFT law given by 
\begin{equation}
    \tau_\alpha =\tau_0\exp\left[ \frac{D}{\phi_0/\phi - 1}\right],
    \label{VFT}
\end{equation} where $D$ is a material dependent parameter, and $\phi_0$ is the cell packing fraction at which $\tau_\alpha$ is expected to diverge. 
In contrast, the dependence of $\tau_s$ is best described by $\tau_s = \tau_0\exp(A\phi)$ (Fig.~\ref{fittedTau} (b)).

\section{Free Volume theory}
\begin{figure}[!htpb]
\begin{center}
\includegraphics[width = 0.95\columnwidth]{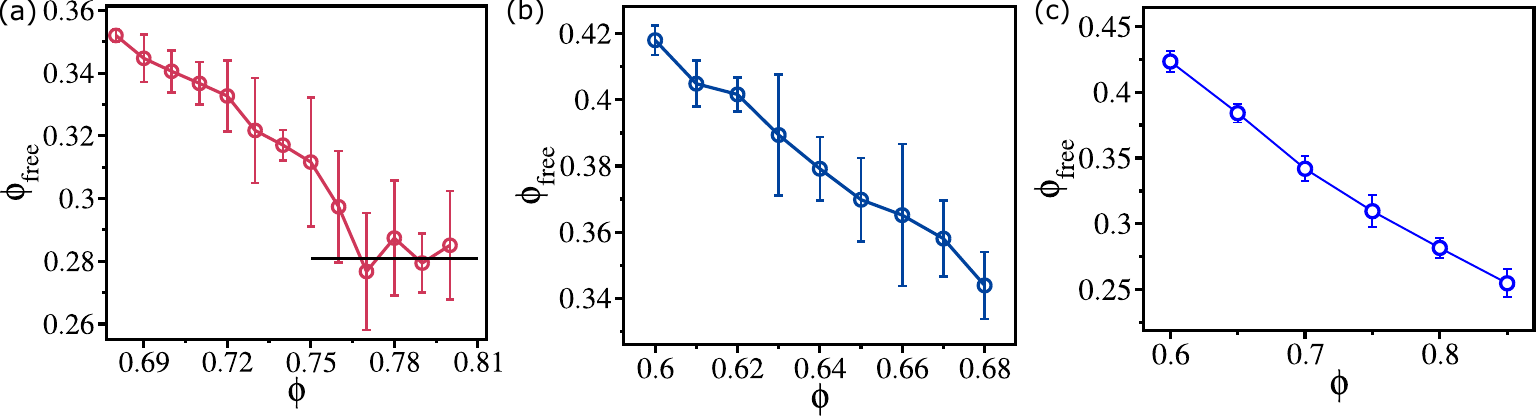}
\caption{\textit{Free volume in 3D tissues:} (a) Free volume $\phi_{\text{free}}$ (Eqn.~\eqref{phifree}) for $E = 0.8E_0$ and $\Sigma = 8.5\%$ (VS regime) as a function of $\phi$. (b) Same as (a), except it is for $E = 5E_0$ (glassy regime). (c) Same as (a) and (b) but for $E = 0.2E_0$ (liquid phase).}
\label{FV}
\end{center}
\end{figure}
In a previous study~\cite{Das2023}, using 2D simulations, we  showed that the saturation of the effective viscosity at high area fraction could be understood using the free volume theory~\cite{CohenJCP1959,Turnbull1961JCP,Turnbull1970JCP,Falk2020PRL}.    In order to determine whether the free volume ($\phi_{\textbf{free}}$) theory explains both the monotonic growth of $\eta_{\text{eff}}$ in the glass-like state as well as the saturation in the VS regime in 3D tissues, we calculated the dependence of free volume fraction $\phi_{\textbf{free}}$ (defined below) as a function of $\phi$.   We estimated   $\phi_{\textbf{free}}$ by first calculating the Voronoi cell volume ($V$) for all the cells at each snapshot. We define $V_{\text{free},i} = V - 4/3\pi R_i^3$ where $R_i$ is the radius of the $i^{th}$ cell. The value of $V_{\text{free}}$ could be negative if the overlap between the neighboring cells is substantial; $V_{\text{free}}$ is positive only when the Voronoi volume is greater than the actual cell volume. The positive value of $V_{\text{free}}$ is an estimate for the available free volume. The availability of free volume for a cell facilitates the mobility of jammed cells by cooperative motion of the neighboring cells. 

The effective free volume fraction $\phi_{\textbf{free}}$ is,
\begin{equation}
\phi_{\text{free}} = \frac{\sum_{j=1}^{N_t}\sum_{i=1}^{N_p} V^j_{\text{free}_{+,i}}}{N_t V_{\text{box}}},
\label{phifree}
\end{equation}
where $N_p$ is the number of cells for which the associated free volume is positive, $V^j_{\text{free+}}$ is the value in $j^{th}$ snapshot, $N_t$ is the total number of snapshots, and $V_{\text{box}}$ is the volume of the simulation box. The effective free volume of certain cells with substantial overlap would have negative values of $V_{\text{free}}$. In such cases, the Voronoi volume of the cell would be less than the actual cell volume. For our purpose, the free volume is the positive value of $V_{\text{free}}$. Only if  $V_{\text{free}}$ is greater than zero, there is available space for the cells to move.  

We find that $\phi_{\textbf{free}}$ saturates when $\phi > \phi_S$ in the VS regime (Fig.~\ref{FV} (a)) whereas it monotonically decreases in the glass-like state (Fig.~\ref{FV} (b)) and liquid (Fig.~\ref{FV} (c)). Interestingly, $\phi_{\textbf{free}}$ decreases almost linearly as $\phi$ increases in the liquid phase, which means that even in the dense ergodic \textcolor{black}{tissue}, the cells can be jammed.  The results show that the notion of free volume provides a unified explanation for both the monotonic growth of $\eta_{\text{eff}}$ in the glass-like state and saturation in the VS. 
\begin{figure}[!htpb]
\begin{center}
\includegraphics[width = 0.95\columnwidth]{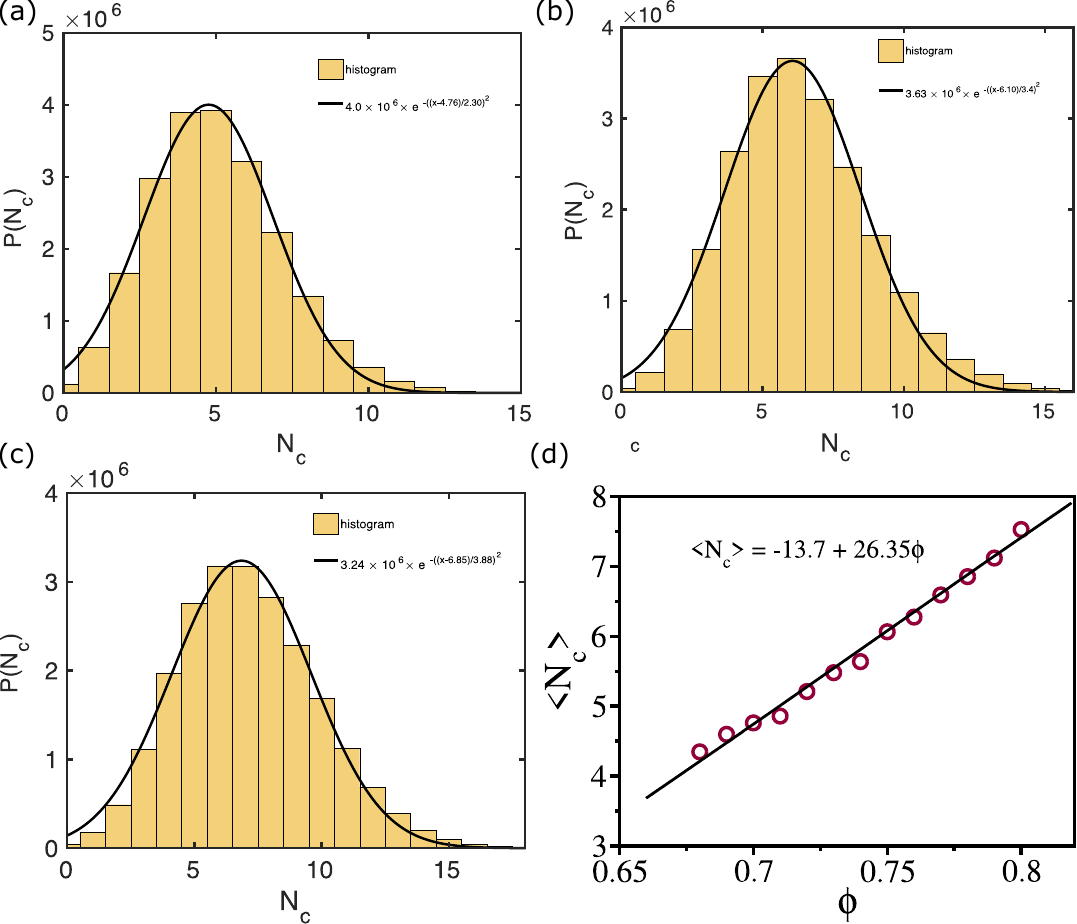}
\caption{\textit{Mean coordination number and cell packing fraction:} $P(N_c)$ at $E = 0.8E_0$ and $\Sigma \sim 8.5\%$ for $\phi = 0.70$ (a), for $\phi = 0.75$ (b) and for $\phi = 0.78$ (c). The solid lines are the Gaussian fits. (d) The mean coordination number $\left\langle N_c \right\rangle$ as a function of $\phi$. The solid line is a linear fit to the data.}
\label{coordination}
\end{center}
\end{figure}

\section{Viscosity depends on cell connectivity}
It is surprising that the striking experimental observation that saturation of viscosity at high $\phi$  in zebrafish blastoderm tissue~\cite{PETRIDOU20211914} could be quantitatively explained using cell-cell contact topology, a purely geometric measure, which was estimated using the average connectivity between cells,  $\left\langle C \right\rangle$. It was also shown that cell packing fraction was linearly related to $\left\langle C \right\rangle$. Two-dimensional simulations~\cite{Das2023} showed that the average coordination number $\left\langle N_c \right\rangle$, which is equivalent to $\left\langle C \right\rangle$ and $\left\langle C \right\rangle$ and $\left\langle N_c \right\rangle$, was linearly related to $\phi$.  More importantly,  we established~\cite{Das2023} that $\eta_{\text{eff}}$ is determined by  $\left\langle N_c \right\rangle$, which was first pointed out in the elsewhere~\cite{PETRIDOU20211914}. We tested if these findings also hold in 3D tissues. 

Let us define the coordination number, $\left\langle N_c \right\rangle$, as the number of cells that are in contact with a given cell. Two cells, with indices $i$ and $j$, are deemed to be in contact if $h_{ij} = R_i+R_j - r_{ij} > 0$.
At each value of $\phi$, we calculated $N_c$ for all the cells and corresponding  distribution $P(N_c)$ (Fig.~\ref{coordination} (a), (b) \& (c)). The distributions $P(N_c)$ are well fit by a Gaussian function $A\exp\left[-(\frac{x- \langle x \rangle}{\sigma})^2 \right]$, where $\langle x \rangle$ is the mean and $\sigma$ is the dispersion. The mean $\left\langle N_c \right\rangle$ calculated from the distribution varies linearly with $\phi$ (Fig.~\ref{coordination} (d)), which implies that $\left\langle N_c \right\rangle \simeq \left\langle C \right\rangle $.
\begin{figure}[!htpb]
\begin{center}
\includegraphics[width = 0.99\columnwidth]{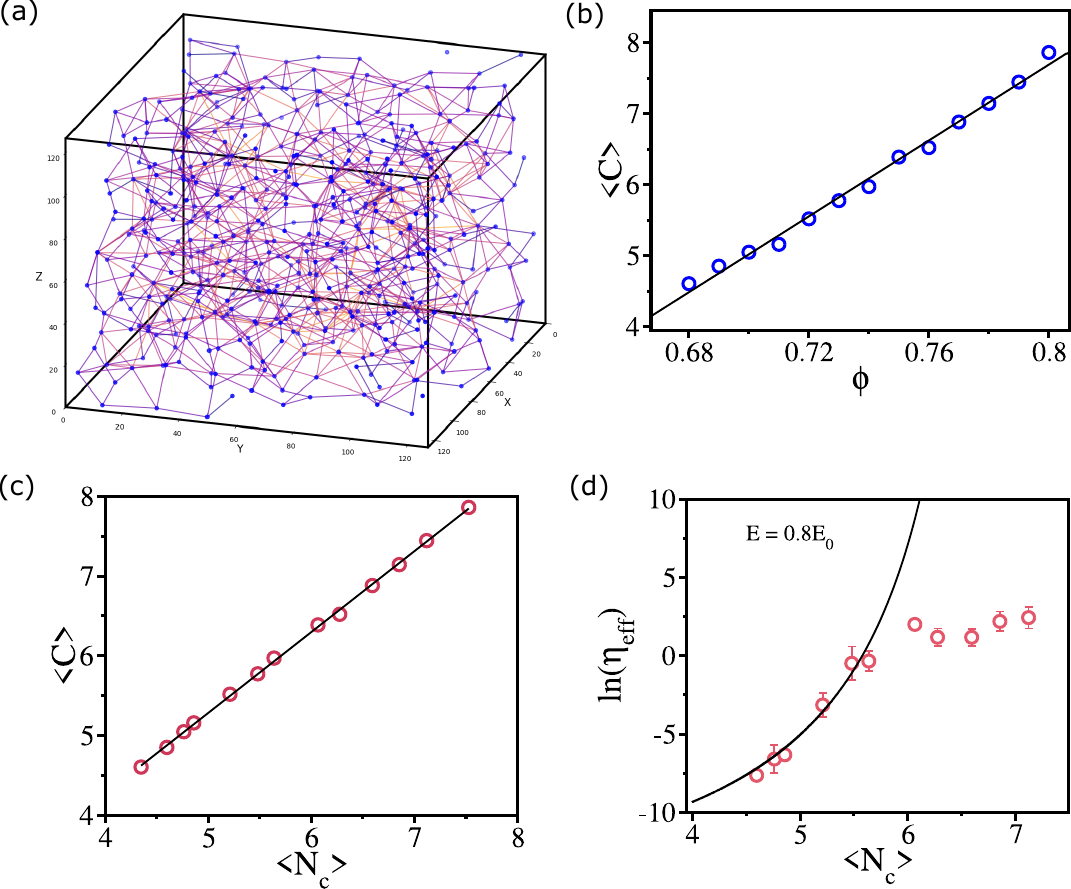}
\caption{\textit{Viscosity and cell contact network topology are related:} (a) Sample connectivity map at $\phi=0.75$, $E=E_0$ and $\Sigma =8.5\%$, corresponding to the VS regime (Fig. 9 (g) in the main text). The blue dots are the cell centers (node) and lines are the edges. (b) Cell connectivity $\left\langle C \right\rangle$ is a linear function of $\phi$.  (c) $\left\langle C \right\rangle$ as a function of network coordination number, $\left\langle N_c \right\rangle$. The solid line is a linear fit to the data. (d) Logarithm of the effective viscosity $\eta_{\text{eff}}$ as a function of $\langle N_c \rangle$. The solid line is the VFT fit (Eqn.~\eqref{VFT}) to the $\eta_{\text{eff}}$ data. }
\label{connectivity}
\end{center}
\end{figure}

Next, we calculated the average connectivity $\left\langle C \right\rangle$. We defined each cell as a node and the line connecting the two nodes as an edge (see Fig.~\ref{connectivity} (a)). If a snapshot has $n$ nodes and $m$ edges then, $\left\langle C \right\rangle = \frac{2m}{n}$. Fig.~\ref{connectivity} (b) shows that $\left\langle C \right\rangle$ is linearly related to $\phi$. Strikingly, $\left\langle C \right\rangle$ and $\left\langle N_c \right\rangle$ has similar values and also are linearly related (Fig.~\ref{connectivity} (c)). Finally,  $\eta_{\text{eff}}$ grows as function of $\left\langle N_c \right\rangle$ (Fig.~\ref{connectivity} (d)), which shows that $\eta_{\text{eff}}$ mirrors the dependence $\left\langle N_c \right\rangle$ on $\phi$. These results show that even in 3D the complex cell contact network determines the viscosity, which is surprising but supports the finding in the previous study \cite{PETRIDOU20211914}.

\section{ On the importance of self-propulsion}
\begin{figure}[!htpb]
\begin{center}
\vskip -0.2cm
\includegraphics[width = 0.95\columnwidth]{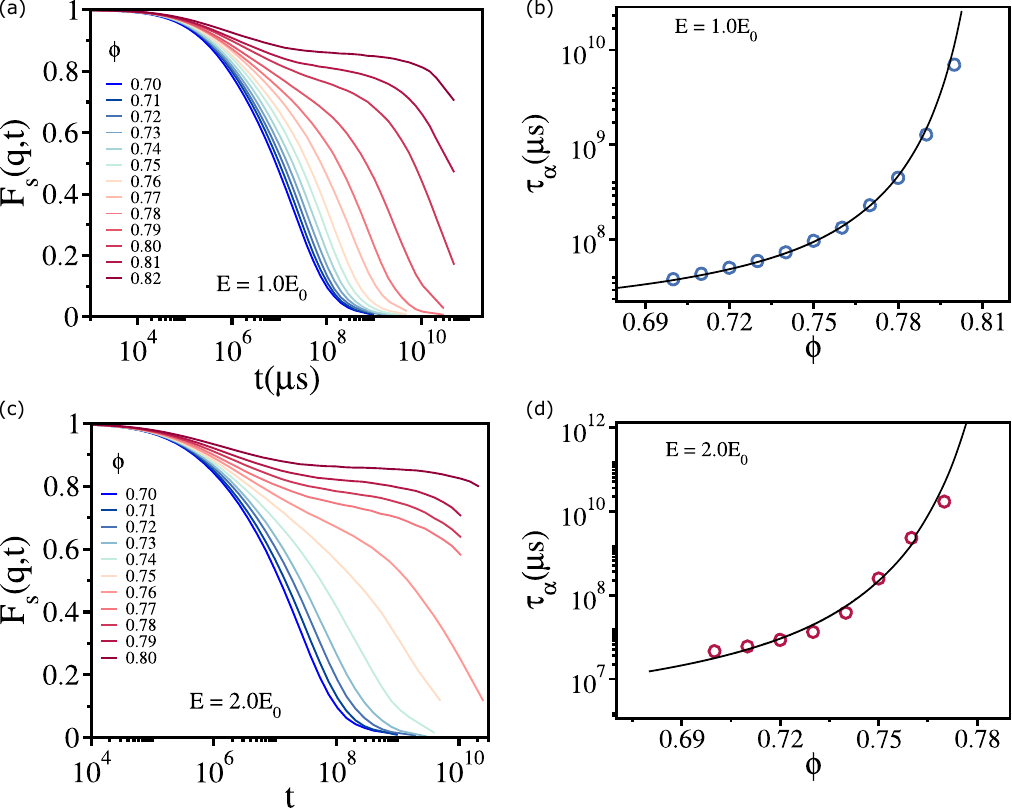}
\caption{\textit{Glassy state in 3D tissues:} (a) $F_s(q,t)$ as a function of $t$ for $0.70 \leq \phi \leq 0.82$. (b) Relaxation time $\tau_\alpha$ as a function  $\phi$.  The solid line is the VFT fit (Eqn.~\ref{VFT}) to the data. We set $\Sigma = 24.8\%$, $E=E_0= 0.001$ MPa and temperature $T = 300K$. (c) Same as (a) except it is for $E = 2.0E_0$ and $\Sigma = 8.5\%$. (d) Same as (b) except for the parameters values, shown in (c).}
\label{Fig5}
\end{center}
\end{figure}
In the main text, we showed that the minimal model that reproduces the saturation in viscosity should include self-propulsion (Eqn. 12 in the main text).
Depending on the parameters ($\Sigma$ and $E$), the tissues exhibit a range of behavior, including the increase in viscosity like in glasses below a critical volume fraction followed by a saturation (the VS regime -- Fig. 9 (g) yellow shaded regime in the main text).  It is natural to wonder if the VS behavior could be reproduced using  simulations performed at a fixed temperature. To explore if the results could be explained by merely varying the temperature, we performed Brownian dynamics simulations  using  the cell model by integrating the equations of motion, 
\begin{equation}
    \dot{\Vec{r_i}} = \frac{\Vec{F_i}}{\gamma_i} + \sqrt{(2k_BT/\gamma_i)}\Vec{\zeta_i}(t),
    \label{BD}
\end{equation}
where $\gamma_i = 6\pi\eta R_i$ with $R_i$ the radius of cell $i$, $\Vec{r_i}(t)$ is the position of a cell at time $t$;  $\Vec{\zeta_i}$ is the random noise that satisfies $\langle \Vec{\zeta_i}(t) \rangle =0$ and $\langle \Vec{\zeta_i}(t) \Vec{\zeta_j}(t') \rangle = 6\frac{k_BT}{\gamma} \delta_{ij} \delta(t-t')$, where $\delta_{ij}$ and $\delta(t-t')$ are Kronecker $\delta$ and Dirac $\delta$ function. Note that unlike the equation of motion used in the main text, Eqn. \ref{BD} obeys the fluctuation-dissipation theorem (FDT).  
\begin{figure}[!htpb]
\begin{center}
\vskip -0.2cm
\includegraphics[width = 0.95\columnwidth]{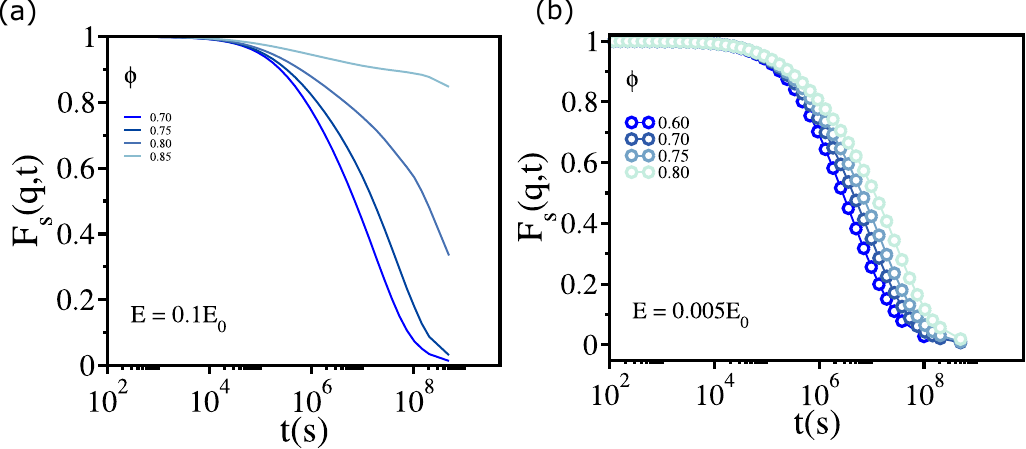}
\caption{\textit{Liquid state in 3D tissues using FDT:} (a) $F_s(q,t)$ as a function of $t$ for $0.70 \leq \phi \leq 0.85$ with $E = 0.1E_0$. (b) Same as (a) but for $E = 0.005E_0$. $\Sigma =8.5\%$ in both (a) and (b).} 
\label{FDTLiq}
\end{center}
\end{figure}
The simulations were performed using $\eta = 5\times10^{-6}$ $kg/(\mu m \cdot s)$, $E = 0.001$ MPa and $\delta t = 10$ $\mu s$. 
For the parameters used in the temperature simulations, we find that the tissue exhibits only the characteristics of glassy dynamics without signs of saturation in $\tau_\alpha$, which is found in Fig.~2 (b), Fig.~2 (d) and Fig.~2 (e) in the main text. The decay  $F_s(q,t)$ occurs in two steps, at most of the values of $\phi$, with a clear plateau, especially at high $\phi$ (Fig.~\ref{Fig5} (a)). The calculated relaxation time $\tau_\alpha$ follows the VFT law (Fig.~\ref{Fig5} (b)). We also performed simulations with $E = 2.0E_0$ and $\Sigma =8.5\%$ and found a similar result (Fig.~\ref{Fig5} (c) and (d)). 

The absence of VS under conditions that satisfy the FDT  does not prove that saturation of $\tau_\alpha$ cannot arise using Eqn. \eqref{BD} if a broader range of cell softness and $\Sigma$ are explored.  However, it is unlikely that the observed VS in the non-equilibrium simulations, with  $(0.3 \lesssim E/E_0 \lesssim 2)$  and $\Sigma > 8\%$ explored in Fig. 7 in the main text, could be reproduced using the FDT-satisfying Brownian dynamics simulations (Eqn. \eqref{BD}).  We surmise that a balance between self-propulsion, $E$, and dispersion in cell size is required to recapitulate the VS behavior.  

We also tested how the liquid state is affected when the self-propulsion is absent. We found that in the presence of finite temperature and in absence of self-propulsion, we need to simulate much softer cells ($E =0.005E_0$) to observe the liquid-like behavior (Fig.~\ref{FDTLiq} (b)), which was found for $E= 0.1E_0$ in the presence of self-propulsion. For temperature simulation with $E = 0.1E_0$, we see a glassy behavior (Fig.~\ref{FDTLiq} (a)). 

\begin{figure}[!htpb]
\begin{center}
\vskip -0.2cm
\includegraphics[width = 0.95\columnwidth]{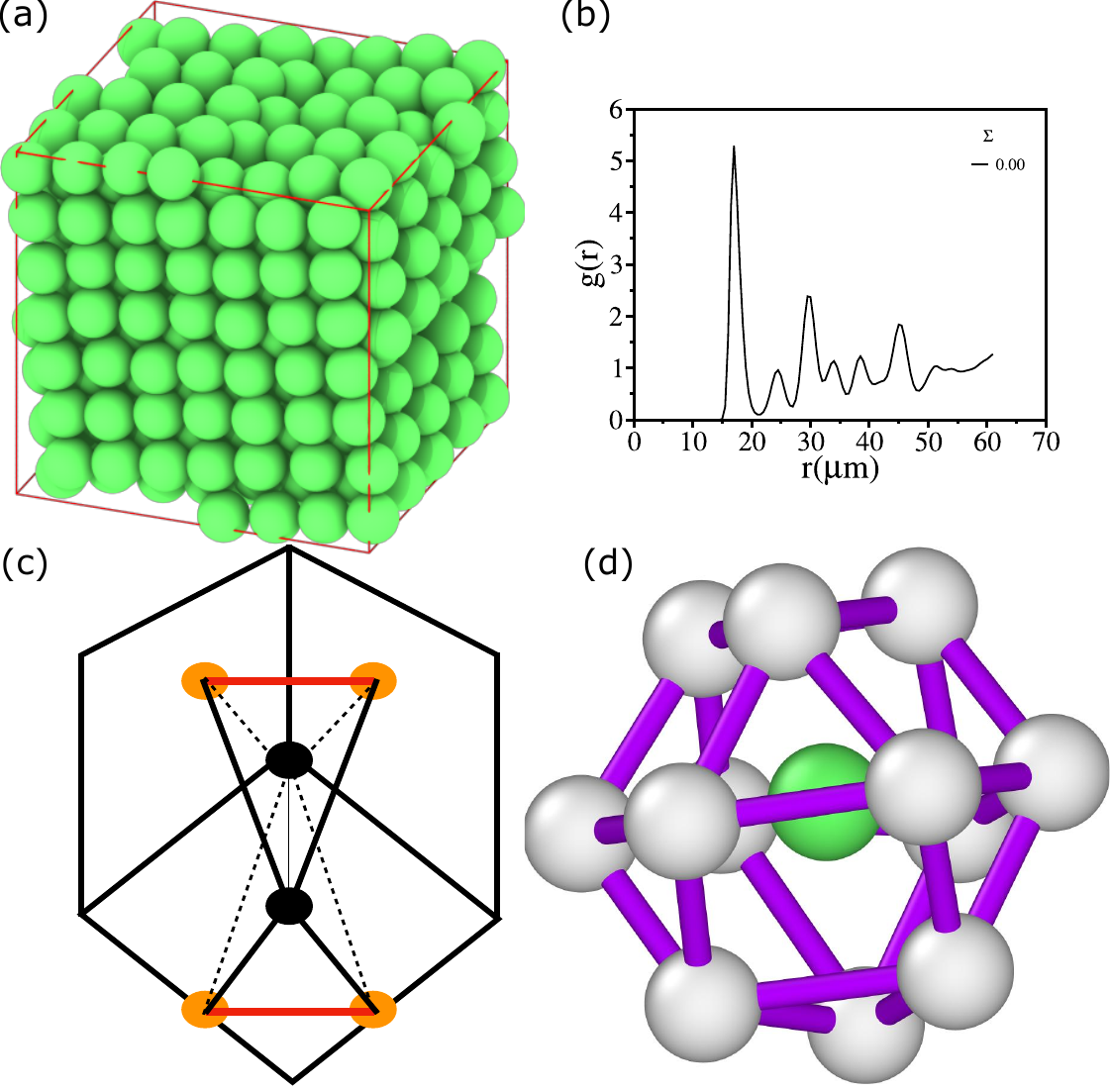}
\caption{\textit{Perfect FCC Crystal using a-CNA:} (a) Simulation snapshot for $E= E_0$ with $\Sigma =0$. The cells are colored green if they match the CNA values for a perfect FCC crystal (4,2,1). (b) The corresponding $g(r)$ as a function of $r$. (c) Illustration of Common neighbor analysis using three indices ($n_{cn}, n_b, n_{1cb}$) for a FCC crystal~\cite{Faken1994}. The bonded atoms in the pair are colored black. The number of neighbors ($n_{cn}$) common to these atoms is colored yellow. The bonds between the common neighbors ($n_b$) are colored in red. (d) Example of total $12$ common neighbors for an FCC crystal type cell (colored green). The actual cell size is scaled to $60\%$ for a better visualization. $E = E_0$ and $\Sigma =0$ and $\phi = 0.77$ for (d).} 
\label{FCC}
\end{center}
\end{figure}
\section{Common Neighbor analysis for identifying crystal structures}
We used the Common Neighbor Analysis (CNA)~\cite{Stukowski2012} method to identify the local atomic structures based on the topology of atomic bonds. In this approach, two atoms are considered bonded if their separation distance is within a predefined cutoff radius ($r_{cut}$), chosen based on the crystal structure. For the FCC and HCP structures,  $r_{cut}$ is set midway between the first and second nearest-neighbor shells, and for BCC structures, it includes both first- and second-nearest neighbors (see Fig.~\ref{FCC} (c) and (d) for illustration). For each bonded pair, a characteristic triplet ($n_{cn}, n_b, n_{1cb}$) is computed,  where $n_{cn}$ represents the number of shared neighbors between two atoms. $n_b$ is the number of bonds between these common neighbors and $n_{1cb}$ is the longest chain of bonds connecting them. The triplets are calculated for a given snapshot and compared with the known values of perfect crystals. For example, the CNA for FCC is (4,2,1), for BCC (6,6,6) for diamond cubic (5,4,3). The CNA method efficiently distinguishes between different crystalline structures. 

In a multiphase system an adaptive common neighbor analysis (a-CNA) was chosen to identify the local atomic structure~\cite{Stukowski2012}. Unlike conventional CNA, which relies on a fixed global cutoff radius, a-CNA determines an adaptive cutoff radius that is specific to each atom based on the local environment. The method begins by identifying the $N_{max}$ nearest neighbors for each atom and sorting them according to their distances. To assess whether the local environment matches a given reference structure, the first $N$ nearest neighbors required for that structure are selected, and a local cutoff radius is computed as a function of their average bond lengths. For FCC and HCP structures, the local cutoff radius is determined using the first 12 nearest neighbors, while for BCC structures, both first and the second nearest neighbors cells are considered.   The calculated values of the triplets ($n_{cn}, n_b, n_{1cb}$) are  compared with the known values for a perfect crystals. The adaptive cutoff improves the  classification accuracy in systems with multiple phases or varying local densities by ensuring that each atom is analyzed using a cutoff that best matches the local immediate environment.
In Fig.~\ref{FCC} we show a snapshot of simulations with $\Sigma=0$ and $E=E_0$ that results in the formation of a perfect FCC crystal.
\section{Two-dimensional simulations}

In the main text, we showed that a 3D non-confluent tissue exhibits the characteristics of a fluid when the $E \lesssim 0.2E_0$, and it exhibits glassy dynamics when $E \gtrsim 3.0E_0 $. These characteristics are found provided the cell size dispersion  $\Sigma$ exceeds a critical value (see the diagram of states in Fig. 7 (d) in the main text).  Additionally, there is a range of $E$ over which the tissue exhibits viscosity saturation as a function of $\phi$. To verify whether a 2D tissue exhibits a similar range of behavior, we performed simulations by varying $E$ at a fixed value of $\Sigma = 24.8\%$, which the estimated previously \cite{Das2023} by analyzing the experimental data \cite{PETRIDOU20211914}. 
\begin{figure}[!htpb]
\begin{center}
\includegraphics[width = 0.95\columnwidth]{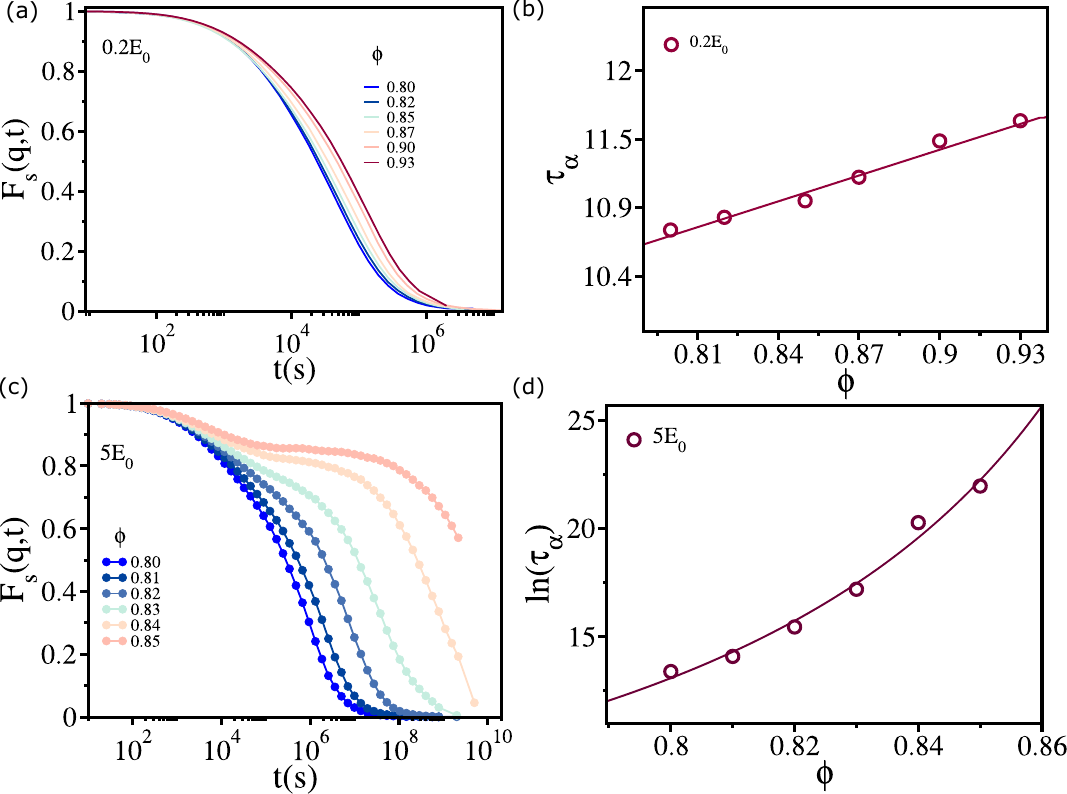}
\caption{\textit{Liquid and glass states in 2D tissues:}  (a) $F_s(q,t)$ as a function of $t$ for various values of $\phi$ with $E = 0.2E_0$ and $\Sigma = 24.8\%$. (b) Relaxation time, $\tau_\alpha$, as a function of $\phi$ for $E = 0.2E_0$ and $\Sigma = 24.8\%$. The solid line is fit to $\tau_\alpha = \tau_0\exp(A\phi)$. (c) Same as  (a) except it is for $E = 5E_0$. (d) Same as (b) but for $E = 5E_0$. The solid line is the VFT fit (see Eqn.~\eqref{VFT}).}
\label{Fig2}
\end{center}
\end{figure}
\begin{figure}[!htpb]
\begin{center}
\includegraphics[width = 0.9\columnwidth]{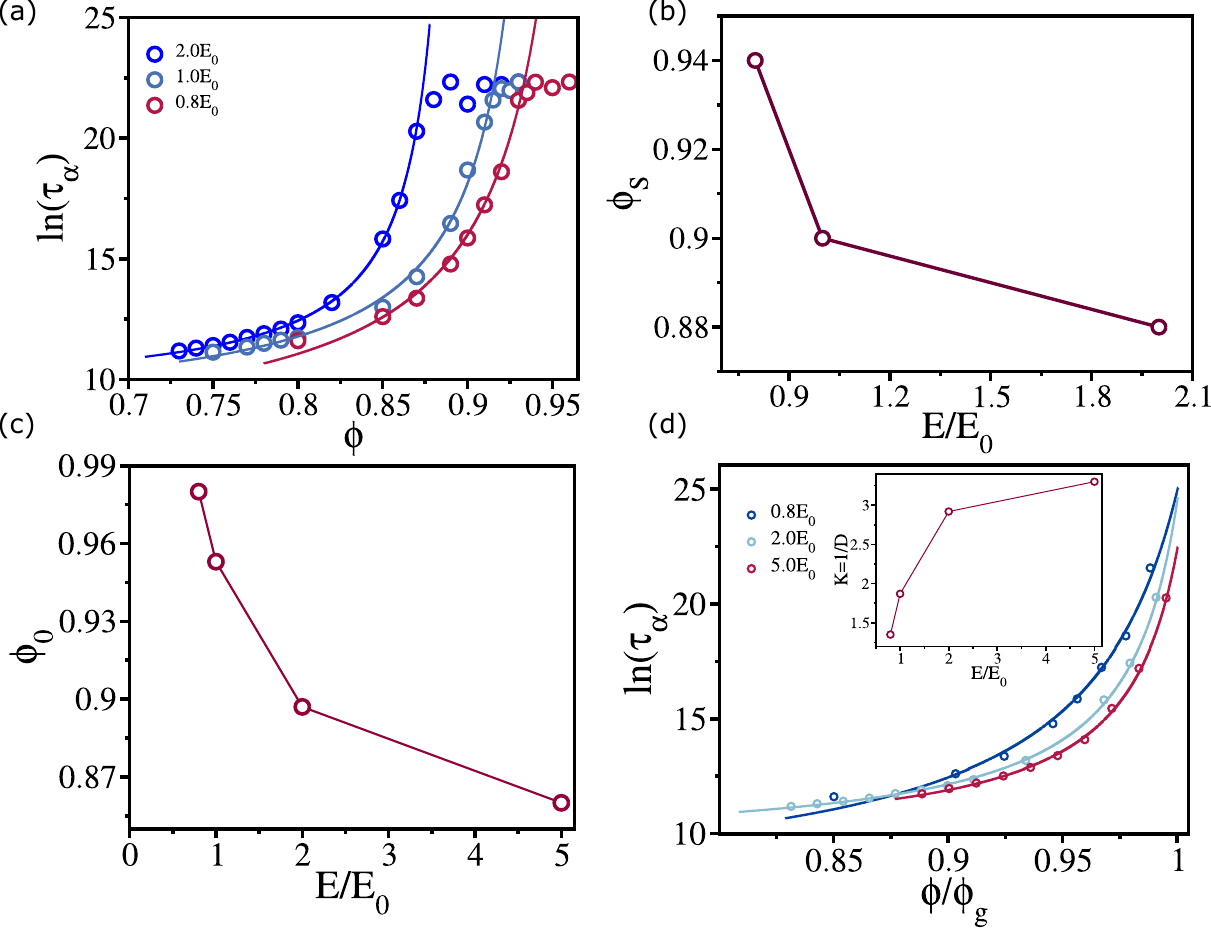}
\caption{\textit{Saturation of $\tau_\alpha$ in 2D tissues:} (a) $\tau_\alpha$ as a function of $\phi$ for $E = 0.8E_0, 1.0E_0$ and $2.0E_0$ with $\Sigma = 24.8\%$. The solid lines are fit to the equation~\eqref{VFT}. (b) The critical packing fraction $\phi_S$ above which $\tau_\alpha$ saturates, as a function of $E$. (c) $\phi_0$ as a function of $E$. (d) $\ln(\tau_\alpha)$ as a function of $\phi/\phi_g$. Inset shows kinetic fragility $K_{VFT} = 1/D$ as a function of $E/E_0$. Here $\Sigma = 24.8\%$. }
\label{Fig1}
\end{center}
\end{figure}

\textbf{Liquid and glassy behavior: } For $E = 0.2E_0$ the decay of $F_s(q,t)$ is nearly independent of $\phi$ for $0.85 \leq \phi \leq 0.93$ (Fig.~\ref{Fig2} (a)), which is reflected in the modest increase in $\tau_\alpha$ as a function of $\phi$ (Fig.~\ref{Fig2} (b)). The tissue behaves like an ergodic liquid. On the other hand, at $E = 5E_0$, the tissue exhibits the characteristics of glass-like dynamics. $F_s(q,t)$ decays in two distinct steps with a clear plateau appearing at the intermediate timescales (Fig.~\ref{Fig2} (c). The duration of the plateau increases as $\phi$ increases. At the intermediate timescales the cells get ``caged'' by their neighbors, which means that the mobility a cell requires movement of a number of other cells. The relaxation time $\tau_\alpha$ increases rapidly and is well described by the VFT law (Eqn.~\eqref{VFT}) (Fig.~\ref{Fig2} (d)).

\textbf{Viscosity saturation (VS):} In the range, $0.8E_0 \leq E \leq 2E_0$, the tissue exhibits the characteristics of viscosity saturation. The relaxation time,  as a function of $\phi$ grows according to the VFT law (Eqn.~\eqref{VFT}) when $\phi \leq \phi_S$. When $\phi$ exceeds $\phi_S$, $\tau_\alpha$ is independent of $\phi$ (Fig.~\ref{Fig1} (a)). As in the 3D tissues, the $\phi_S$ value decreases as $E$ increases (Fig.~\ref{Fig1} (b).
In addition,  $\phi_0$ decreases (Fig.~\ref{Fig1} (c)) and fragility the rate at which $\tau_\alpha$ increases with $\phi$) increases with $E$ (Fig.~\ref{Fig1} (d)). We calculated the kinetic fragility $K = 1/D$ (see Eqn.~\eqref{VFT}). As shown in the inset of Fig.~\ref{Fig1} (d), it increases by a factor of $\sim 2.5$ when $E$ changes from $0.8E_0$ to $5E_0$ which suggests as the cells become more deformable, the glass transition should appear at a higher packing fraction than for an equivalent hard sphere system. \textcolor{black}{In synthetic materials the glass transition packing fraction $\phi_g$ is defined as the packing fraction where viscosity $\sim 10^{13}$ poise or relaxation time goes to $\sim 100$ s. Following the the protocol used in the glass transition literature,  we define $\phi_g$ in our system as the $\phi$ where $\tau_\alpha \simeq 10^{11}$. 
For reference, the glass transition packing fraction $\phi_g \sim 0.58-0.60$~\cite{vanMegen1994PRE} for pure hard-sphere liquids. On the other hand for soft tissues, it can occur at a much higher packing fractions~\cite{Angelini2011,Bi2015}.} 

Taken together, the results in presented in this section show that all the phases and dynamical behavior found in 3D are recapitulated in 2D as well. 
\newpage
\section{Simulation parameters}
\begin{table}[ht!]
\centering 
\caption{Parameters used in the simulations.}
\label{table1}
\scalebox{1}{\begin{tabular}{ |p{4cm}||p{2cm}|p{2cm}|p{2cm}|  }
 \hline
 \hline
 \bf{Parameters} & \bf{Values} & \bf{References} \\
 \hline
 \hline
 Timestep ($\Delta t$)& 10$ \mathrm{s}$  & \textcolor{black}{\cite{Das2023}} \\
 \hline
Self-propulsion ($\mu$) &  0.045 $\mathrm{\mu m/\sqrt{s}}$ & \textcolor{black}{\cite{Das2023}} \\
 \hline
Friction coefficient ($\gamma_o$) & 0.1 $\mathrm{kg/ (\mu m~s)}$   & \textcolor{black}{\cite{Das2023}}  \\
 \hline
Mean Cell Elastic Modulus ($E_{i}) $ & $0.1E_0 - 8.0E_0$  & ~\cite{galle2005modeling,malmi2018cell}    \\
 \hline
Mean Cell Poisson Ratio ($\nu_{i}$) & 0.5 & ~\cite{schaller2005multicellular,malmi2018cell}  \\
 \hline
\hline
\end{tabular}}
\end{table}
\newpage
\bibliography{3DCell}
\bibliographystyle{unsrt}